\documentclass[aps,prx,twocolumn,notitlepage,showpacs,amsmath,amstex,amssymb,citeautoscript,longbibliography]{revtex4-1}

\pdfoutput=1
\usepackage{natbib}
\usepackage[english]{babel}
\usepackage{letltxmacro}
\usepackage{latexsym}
\usepackage{tabularx,tabu}

\LetLtxMacro{\ORIGselectlanguage}{\selectlanguage}
\makeatletter
\DeclareRobustCommand{\selectlanguage}[1]{%
  \@ifundefined{alias@\string#1}
    {\ORIGselectlanguage{#1}}
    {\begingroup\edef\x{\endgroup
       \noexpand\ORIGselectlanguage{\@nameuse{alias@#1}}}\x}%
}
\newcommand{\definelanguagealias}[2]{%
  \@namedef{alias@#1}{#2}%
}
\makeatother
\definelanguagealias{en}{english}
\definelanguagealias{English}{english}
\usepackage{graphicx}
\usepackage{amsmath}
\usepackage{amsfonts}
\usepackage{amssymb,bbding}
\usepackage{bm}
\usepackage{color}
\usepackage[percent]{overpic}
\usepackage{soul} 
\usepackage{amssymb}
\usepackage{wasysym}
\usepackage{dsfont}
\usepackage{float}
\usepackage{braket}
\usepackage{cancel}
\usepackage{comment}
\usepackage{hyperref}
\usepackage{tikz-feynman,subcaption}
\usepackage{ragged2e}
\DeclareCaptionJustification{justified}{\justifying}
\usepackage[font=small,labelfont=bf,
   justification=justified,
   format=plain]{caption}

\makeatletter
\newcommand{\thickhline}{%
    \noalign {\ifnum 0=`}\fi \hrule height 1.2pt
    \futurelet \reserved@a \@xhline
}
\newcolumntype{"}{@{\hskip\tabcolsep\vrule width 1.2pt\hskip\tabcolsep}}
\makeatother

\renewcommand{\Im}[0]{\operatorname{Im}}

\begin{document}

\title{Corrections to diffusion in interacting quantum systems}

\author{Alexios A.~Michailidis}
\affiliation{\em Department of Theoretical Physics, Universit\'e de Gen\`eve,
 1211 Gen\`eve, Switzerland}
 \affiliation{PlanQC GmbH, Lichtenbergstr. 8, 85748 Garching, Germany}
\author{Dmitry A.~Abanin}
\affiliation{\em Department of Theoretical Physics, Universit\'e de Gen\`eve,
 1211 Gen\`eve, Switzerland}
 \affiliation{Department of Physics, Princeton University, Princeton, New Jersey 08544, USA}
\author{Luca V. Delacr\'etaz}
\affiliation{\em Department of Theoretical Physics, Universit\'e de Gen\`eve,
 1211 Gen\`eve, Switzerland}
\affiliation{James Franck Institute, University of Chicago, Chicago, Illinois 60637, USA}\affiliation{Kadanoff Center for Theoretical Physics, University of Chicago, Chicago, Illinois 60637, USA}

\begin{abstract}

Transport and the approach to equilibrium in interacting classical and quantum systems is a challenging problem of both theoretical and experimental interest. One useful organizing principle characterizing equilibration is the dissipative universality class, the most prevalent one being diffusion. In this paper, we use the effective field theory (EFT) of diffusion to systematically obtain universal power-law corrections to diffusion. We then employ large-scale simulations of classical and quantum systems to explore their validity. In particular, we find universal scaling functions for the corrections to the dynamical structure factor $\langle n(x,t)n\rangle$, in the presence of a single $U(1)$ or $SU(2)$ charge in systems with and without particle-hole symmetry, and present the framework to generalize the calculation to multiple charges.  Classical simulations show remarkable agreement with EFT predictions for subleading corrections, pushing precision tests of effective theories for thermalizing systems to an unprecedented level. Moving to quantum systems, we perform large-scale tensor-network simulations in unitary and noisy 1d Floquet systems with conserved magnetization. We find a qualitative agreement with EFT which becomes quantitative in the case of noisy systems. Additionally, we show how the knowledge of EFT corrections allows for fitting methods, which can improve the estimation of transport parameters at the intermediate times accessible by simulations and experiments. Finally, we explore non-linear response in quantum systems and find that EFT provides an accurate prediction for its behavior.  
Our results provide a basis for a better understanding of the non-linear phenomena present in thermalizing systems.

\end{abstract}
\maketitle
\section{Introduction}
One of the main pursuits of condensed matter physics is the understanding of out-of-equilibrium phenomena in many-body systems. Transport probing the only slightly out-of-equilibrium, linear response, regime is particularly accessible experimentally and therefore most important to understand theoretically, to tie experiments with insight into the fundamental structure of correlated matter.The experimental accessibility of linear response observables has allowed to establish some of the most puzzling phenomenology in condensed matter physics, including the $T$-linear resistivity \cite{PhysRevLett.59.1337} of high-$T_c$ superconductors and heavy fermion systems, anomalous Hall angles \cite{PhysRevLett.67.2088} and magnetoresistance \cite{hayes2016scaling}, which have largely eluded explanations despite decades of activity. More recently, experiments in synthetic quantum matter, such as cold atoms and superconducting quantum circuits, have offered new tools to explore quantum transport (e.g., \cite{Schneider2012,schreiber2015observation,wei2022quantum,morvan2022formation,mi2023stable}), further emphasizing the need for a better theoretical understanding of the landscape of transport phenomena in many-body systems. The theoretical challenge lies in finding controlled methods to study dynamics in strongly correlated systems.

Hydrodynamics -- broadly understood as the emergent dynamics of conserved densities in thermalizing systems -- offers particularly suitable tools in this regard, providing a framework to parametrize and understand near-equilibrium dynamics at late times. In the hydrodynamic limit, the dynamics typically follow a universal behavior, characterized by a dissipative universality class, the most prevalent being diffusion. While deriving the hydrodynamics of diffusion from microscopics  is typically a challenging task, experimental and numerical evidence strongly suggests that it describes the leading order linear response at late times of thermalizing classical and quantum many-body systems, across diverse scales and platforms.Yet, the dissipative universality classes provide information  beyond the leading late-time behavior of linear response: they also include nonlinear response, and universal scaling corrections to observables. In particular, the leading late time behavior of simple observables such as the dynamic structure factor $\langle n(x,t)n\rangle$ can be found from classical hydrodynamic equations \cite{kadanoff1963hydrodynamic}. However, the understanding of corrections to linear response and more complicated observables requires a framework for hydrodynamic fluctuations that systematically treats noise. Several proposals for doing so exist, including generalizations of the Martin-Siggia-Rose formalism \cite{Martin:1973zz} to allow for non-Gaussian noise, Fokker-Planck equations for continuous fields (e.g., \cite{PhysRevLett.127.072301}), macroscopic fluctuation theory \cite{RevModPhys.87.593}, and effective field theories on Schwinger-Keldysh contours \cite{Crossley:2015evo}. It is not clear which of these effective theories -- if any -- describes thermalizing many-body systems beyond the leading late time behavior. Furthermore, the differences between classical and quantum systems in terms of hydrodynamic fluctuations remain ambiguous, as well as the capacity of these effective theories to discern them.

\begin{figure}
	\centering
	\includegraphics[width=\columnwidth]{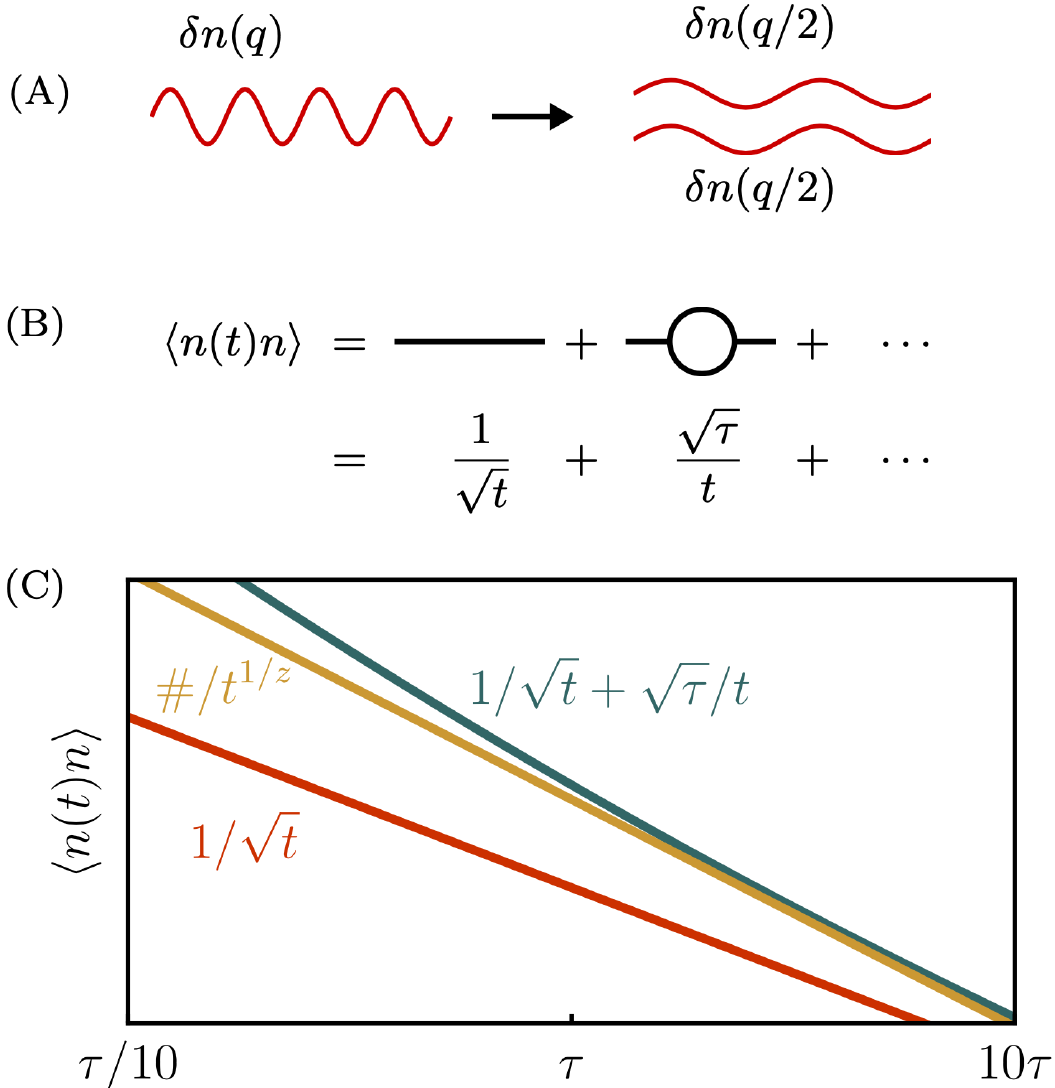}
 \caption{\label{Fig:1} (A) Nonlinear fluctuations of conserved densities are present in generic many-body systems, and (B) lead to universal corrections to hydrodynamics at late times. (C) In the case of a single diffusive density, the leading correction is positive, and can cause a diffusive system (with autocorrelation function illustrated in green) to appear superdiffusive (yellow, $z=3/2$ is shown above) at intermediate times. The EFT of diffusion predicts the coefficient of this correction $\tau = \frac{\chi^2 D'^2}{16\pi D^4}$, together with a universal scaling function of $x/\sqrt{Dt}$, see Eqs.~\eqref{eq_nn_noph} and \eqref{eq_F10}.}
\end{figure}

Beyond identifying the correct theory of fluctuations, understanding corrections to observables in thermalizing systems has important experimental and numerical implications. 
Starting with numerics, a systematic theory of scaling corrections is critical for quantum simulations, which can typically access intermediate times, during which the effects of corrections can be significant. This can result in inaccurate determination of transport parameters, or even in an incorrect value of the dynamical exponent $z$, as illustrated in Figure~\ref{Fig:1}. Additionally, diffusive dynamics are also present in non-generic systems, such as certain integrable systems~\citep{Doyon18PRL,DoyonSP19,Vasseur18PRB} and non-interacting systems where diffusion is induced by noise. In these cases, even if the leading late time behavior is the same, the scaling corrections are sensitive to the number and type of conserved densities; they therefore offer precision tests of thermalization, unambiguously distinguishing various apparently diffusive systems. In experiments, the presence of these corrections has interesting consequences for the understanding of thermalization in correlated materials. Power-law corrections to late-time observables come with time scales related to the local equilibration time -- the time scale at which regular hydrodynamics kicks in. This time scale is parametrically large in weakly coupled or nearly integrable systems, but seemingly cannot be made arbitrarily small at strong coupling; this has lead to the expectation that the local equilibration time is universally bounded by the `Planckian' time, $\hbar/T$ \cite{sachdev2007quantum,Zaanen2004,Bruin_Planckian,Hartnoll:2021ydi}. We will show that the leading power-law corrections are in fact entirely fixed by derivatives of diffusivities with respect to the equilibrium value of the transported density or associated potential $D'\equiv dD/dn$ (e.g.~temperature for heat diffusion); hydrodynamics, therefore, universally ties nonlinear response, scaling corrections to linear response, and dependence of transport coefficients on experimental tuning parameters. Since the latter are readily available in experiments and numerics, this provides a time scale that must be exceeded to access the asymptotic regime.

In this paper, we use the effective field theory (EFT) of diffusion \cite{Crossley:2015evo} to systematically and quantitatively study the corrections to observables in generic diffusive systems. The EFT relies on two mild assumptions: (1) the locality of the generator of the dynamics, and (2) the thermalization of the system, i.e. the only collective excitations that survive at late times are conserved densities and associated noise fields. Therefore, it is expected to apply to a broad range of quantum and classical systems. We focus on the dynamic structure factor $\langle n(x,t) n\rangle$ in 1d lattice systems ($>$1d generalizations are shown in Appendix~\ref{ap:A3}) where nonlinear corrections are particularly strong, but also study nonlinear response (in Appendix~\ref{ap:NL}), which offers a complementary  verification of the theory's predictions. The potential significance of scaling corrections to correlation functions in quantum systems has been appreciated for some time~\cite{PhysRevB.73.035113,PhysRevA.89.053608}; however, even the corrections to the considerably simpler autocorrelation function, $\langle n(0,t)n\rangle$, were only obtained analytically and observed numerically recently~\cite{Chen-Lin:2018kfl,Glorioso:2020loc}. 

We start by sharpening these results and generalizing the approach in several directions, to ultimately construct a theory of scaling corrections in thermalizing systems, providing a framework to make quantitative predictions systematically in an expansion at late times. We first show that the coefficient of the leading correction is entirely fixed in terms of transport parameters of the system and their derivatives with respect to equilibrium densities -- this allows us to establish that the correction is non-negative in the case of a single diffusing density, making generic diffusive systems appear superdiffusive at intermediate times (Fig.~\ref{Fig:1}). Next, generalizing to the dynamic structure factor $\langle n(x,t) n\rangle$, we find the universal scaling function of $x/\sqrt{Dt}$ accompanying this correction. We also compute subleading corrections, which arise from higher-order (2-loop) fluctuation effects, as well as higher derivative terms in the EFT. These are particularly important in systems with particle-hole symmetry, where the leading, 1-loop, correction is absent. These new corrections come with their own universal scaling functions of $x/\sqrt{Dt}$, summarized in Table~\ref{tab:1}. We also present the EFT framework required to study study corrections in the presence of multiple diffusive charges and derive the corrections for the case of chaotic spin chains with $SU(2)$ symmetry.

\begin{table}[!t]
\setlength{\tabcolsep}{0.2em} 
{\renewcommand{\arraystretch}{1.7}
\begin{tabular}{ |c"c|c| } 
 \hline \, Leading order\,  & \, no particle-hole \, & particle-hole  \\ 
 \thickhline non-linear  & $\frac{1}{t^{d/2}} F_{1,0}$ & $\frac{1}{t^d} \left(F_{2,0}+\log(t) \tilde F_{2,0}\right)$ \\
 \hline linear & $\frac{1}{t} F_{0,1}$ & $\frac{1}{t} F_{0,1}$    \\ 
 \hline
\end{tabular}
}
\caption{\label{tab:1}Leading order corrections to the dynamical structure factor of chaotic diffusive systems from loop corrections (non-linear) or higher derivative corrections (linear), see Eqs.~(\ref{eq_nn_noph},\ref{eq_nn_ph}). The leading order non-linear correction (1-loop) vanishes in the presence of particle-hole symmetry, and therefore the sub-leading correction (2-loop) dominates. These non-linear corrections are the leading corrections in one dimension ($d = 1$).}
\end{table}

We then quantitatively test these predictions in numerics. We first consider classical lattice gases where $D(n)$ is known analytically, so that the theory prediction can be compared to simulations without requiring any fitting parameter. We find remarkable agreement for the entire scaling function accompanying the correction to diffusion, shown in Fig.~\ref{Fig:4}, thereby providing a test of theories of fluctuating hydrodynamics with unprecedented level of precision.
We next show that the EFT corrections are also present in the dynamics of interacting quantum spin-chains. In this case, the classical resources required to accurately capture the dynamics grow rapidly with the simulation time, and therefore our simulations cannot always reach asymptotic times. We demonstrate that incorporating the EFT corrections into the fitting process leads to considerably more accurate transport parameters, such as diffusivity. 

Finally, we discuss nonlinear response. We show that the EFT universally ties higher-point functions of densities  \cite{Delacretaz:2023ypv} to scaling corrections to linear response. These observables can therefore be used to understand which timescale must be exceeded to enter the asymptotic, late time or low frequency, regime. As controlled experimental probes of nonlinear response improve \cite{Nicoletti:16,chaudhuri2022anomalous}, this offers a quantitative correspondence between these observables and thermalization. We expect these nonlinear observables, as well as fluctuation corrections to linear response, to be within reach of current experiments in cold atoms as well~\cite{Schmiedmayer17Nature,wei2022quantum,mi2023stable}. Measuring higher-point functions in numerics can also help establish unambiguously the dissipative universality class with limited resources.

The paper is organized as follows. In Section~\ref{sec_scalingfunctions} we present the leading corrections to the full dynamical structure factor in 1d, for systems with one abelian local charge. Next we present the corrections for systems which additionally exhibit particle-hole symmetry, and therefore exhibit a vanishing leading correction. These corrections originate from both linear and non-linear fluctuations. However, non-linear fluctuations are logarithmically stronger in 1d, and are expected to dominate at long times. In Section~\ref{sec_EFT} we formulate the EFT formalism and present the main steps towards the calculation of the corrections. In Section~\ref{sec_EFT}.A we derive the leading 1-loop corrections, and in Section~\ref{sec_EFT}.B we outline the basic steps for the 2-loop calculation required to get the non-linear corrections in systems with particle-hole symmetry. Then, in Section~\ref{sec_EFT}.C we discuss the structure of linear corrections. Section~\ref{sec_EFT}.D extends our results to systems with multiple densities. In particular, we present the result for a single non-abelian ($SU(2)$) charge. We conclude by numerically verifying the leading order corrections for classical systems, Section~\ref{sec_EFT}.E. In Section~\ref{sec_QT} we study the linear response regime of quantum systems (coherent and incoherent) with magnetization conservation. For incoherent systems we quantitatively verify the presence of EFT predictions. Coherent dynamics are more complex and they display longer transient phenomena which persist on all simulated timescales. Nevertheless, our results qualitatively agree with EFT. In the concluding Section~\ref{sec_disc}, we consolidate our findings, discuss the relevance of these results for the field, and outline potential avenues for future research. In Appendix~\ref{sec_nonlin} we explore the nonlinear response through a simple three-point function, which offers a complementary test for the validity of EFT.

\section{Full scaling function--analytical predictions}\label{sec_scalingfunctions}

Whenever diffusion or any other hydrodynamic behavior emerges in a many-body system, it is inevitably accompanied by scaling corrections that may be important at intermediate times. These arise from higher-derivative corrections \cite{landau2013fluid} as well as fluctuation (or `loop') corrections \cite{PhysRevA.1.18} in the hydrodynamic description. While these corrections have been appreciated in the context of quantum many-body systems for some time (see, e.g.,~\cite{PhysRevB.73.035113}), they are often 
ignored. Since accessing late times in quantum simulations is fairly prohibitive, accounting for these corrections to scaling is crucial even to correctly capture the dissipative universality class of a given system.

One central result of this work is that these scaling corrections come with entire universal scaling {\em functions}, which can be obtained from the EFT \cite{Crossley:2015evo}. For example, the leading correction to diffusive correlation functions in one dimension comes from a 1-loop correction and takes the form
\begin{equation}\label{eq_nn_noph}
\langle n(x,t)n\rangle
	= \frac{\chi}{\sqrt{4\pi D t}} \left[F_{0,0}(y) + \frac{1}{\sqrt{t}} F_{1,0}(y) + O \left(\frac{\log t}{t}\right)\right] ,
\end{equation}
where $F_{0,0}$ and $F_{1,0}$ are scaling functions of the scaling variable $y\equiv x/\sqrt{Dt}$. The leading scaling function $F_{0,0}(y) = e^{-y^2/4}$ solves the linearized diffusion equation, whereas the leading correction $F_{1,0}(y)$ comes from a 1-loop contribution \cite{Chen-Lin:2018kfl} which, as we show below, takes the form
\begin{align}\label{eq_F10}
F_{1,0}(y)
	&= \frac{\chi D'^2}{D^{5/2}} \tilde F_{1,0}(y)\\ \notag
\tilde F_{1,0}(y)
	&= \frac{4+y^2}{16\sqrt{\pi}} e^{-y^2/2} + \frac{y(y^2-10)}{32} e^{-y^2/4}{\rm Erf}(y/2)
\end{align}
We have separated $F_{1,0}$ into a universal dimensionless function $\tilde F_{1,0}$, and a non-universal factor that involves the susceptibility $\chi$ and the diffusivity $D$ (like the leading order correlator), but also the derivative of the diffusivity with respect to the background value of the diffusing density $D'\equiv dD(n)/dn$. If this parameter is known, e.g.~by measuring the diffusivity at several densities, the entire functional form of the $1/\sqrt{t}$ correction to diffusion is fixed. One interesting feature of this correction is that for $x=0$ it is non-negative $F_{1,0}(0) = \frac{\chi D'^2}{D^{5/2}} \frac{1}{4\sqrt{\pi}}\geq 0$, which implies that the autocorrelation function approaches its asymptotic diffusive form from above at late times:
\begin{equation}\label{eq:autocor}
\langle n(0,t) n\rangle
	= \frac{\chi}{\sqrt{4\pi D t}} \left( 1 + \frac{\chi D'^2}{4\sqrt{\pi} D^{5/2}}\frac{1}{\sqrt{t}} + O\Bigl(\frac{\log t}{t}\Bigr)\right)\, .
\end{equation}
Therefore, if a dynamic critical exponent $z$ is extracted by fitting the autocorrelation function as $\langle n(0,t)n\rangle \sim 1/t^{1/z}$ at late times, a diffusive system will always naively appear to be superdiffusive, $z<2$. This is illustrated in Fig.~\ref{Fig:1}.

Eq.~\eqref{eq_nn_noph} includes the first two terms in a general expansion in derivatives and fluctuations, whose structure is shown in Eq.~\eqref{eq_correlator_generalform}. The correction to diffusion that arises from $\ell$-loop contributions at $n$th order in the derivative expansion in the EFT scales as $1/t^{n+\ell d/2}$ in $d$ spatial dimensions and is encoded in a scaling function $F_{\ell,n}(y)$, which is universal up to one (or a few) non-universal Wilsonian coefficients, similar to the functions in Eq.~\eqref{eq_F10}.

Given that diffusivities generically depend on density, the leading correction \eqref{eq_F10} is typically present. However, $D'$ may vanish at special values of the density: this occurs for example if there is a particle-hole (or charge conjugation) symmetry, which commonly arises in lattice systems at half-filling. In this case, the leading correction to diffusion takes the form
\begin{align}\label{eq_nn_ph}
&\langle n(x,t)n\rangle
	= \frac{\chi}{\sqrt{4\pi D t}}  \biggl[F_{0,0}(y) \\
	& + \frac{1}{t} \left(F_{0,1}(y) + F_{2,0}(y) + F_{2,0}'(y)\log t\right) + O \left(\frac{\log t}{t^2}\right)\biggr], \notag
\end{align}
and comes a higher-derivative contribution $F_{0,1}(y)$, and a two-loop contribution $F_{2,0}(y)+F_{2,0}'(y)\log t$. The former can be shown to take the form
\begin{equation}\label{eq_F01}
\begin{split}
F_{0,1}(y)
	&= \left[c_1 (y^2-2) + c_2y^2(y^2-6)\right] e^{-y^2/4}\, , 
\end{split}
\end{equation}
where $c_1,\, c_2$ are non-universal transport parameters, 
while the latter is obtained in this paper, and is given by:
\begin{equation}\label{eq_F20}
\begin{split}
F_{2,0} &+ F_{2,0}'\log t
	= \frac{\chi^2 D''^2}{12\sqrt{3\pi} D^3} \left[\tilde F_{2,0} + \tilde F_{2,0}'\log t\right]\\
\tilde F_{2,0}(y)
	&= \int_0^\infty \frac{ds}{\pi} \cos (sy) s^2 \\
	&\!\! \times\left[ s^2e^{-s^2} \left(\log \frac{1}{s^2} + {\rm Ei} \left(\tfrac{2s^2}{3} \right)\right)  - \frac32 e^{-s^2/3} \right] \\
\tilde F_{2,0}'(y)
	&= \frac{y^4-12y^2+12}{32 \sqrt{\pi}} e^{-y^2/4}\, , \\
\end{split}
\end{equation}
with  ${\rm Ei}(z)\equiv -\int_{-z}^\infty \frac{du}{u} e^{-u}$. In the first line, we again separated the scaling function into a non-universal factor, which now depends on $D''\equiv d^2 D(n)/dn^2$, and a universal scaling function. Notice that a shift in the logarithm $\log t\to \log (t/\tau)$ can be absorbed by the higher derivative corrections $c_1,\, c_2$ in \eqref{eq_F01}. At asymptotically late times, the diffusive autocorrelation function is again approached from above:
\begin{equation}
\langle n(0,t) n\rangle
	= \frac{\chi}{\sqrt{4\pi D t}} \left( 1 + \frac{\chi^2 D''^2}{32 \sqrt{3}\pi D^3}\frac{\log t}{t} + O\Bigl(\frac{1}{t}\Bigr)\right)\, .
\end{equation}
We note, however, that this correction has only a $\log t$ enhancement compared to the non-sign definite $1/t$ corrections from \eqref{eq_F01}. For reader's convenience, the universal scaling functions found above are illustrated in Fig.~\ref{Fig:scaling_functions} in  Appendix~\ref{ap:A}.

\section{Scaling corrections from the EFT}\label{sec_EFT}

The universal corrections to diffusion quoted in Eqs.~\eqref{eq_nn_noph} and \eqref{eq_nn_ph} can be obtained from the effective field theory (EFT) of diffusion \cite{Crossley:2015evo,Chen-Lin:2018kfl}. These arise from thermal fluctuations (loops) of the hydrodyanmic densities and noise fields. Several qualitative properties of these loops were already understood shortly after their discovery in classical numerics \cite{PhysRevA.1.18}, e.g.~through mode-coupling approximations or the Martin-Siggia-Rose approach \cite{Martin:1973zz}, which established first steps towards a general EFT for fluctuating hydrodynamics. 
The modern EFT approach completes these constructions by elevating them into a systematic expansion in derivatives and fluctuations; we will therefore follow this approach here.

We are interested in studying transport in a system with at least one conserved quantity, leading to a continuity equation (in the continuum limit)
\begin{equation}\label{eq_continuityrel}
\dot n + \nabla \cdot j = 0\, .
\end{equation}
The density $n$ could correspond to energy density, charge density, magnetization density, etc. We will first focus on the situation where a single density is conserved, and discuss generalizations to multiple densities in Sec.~\ref{sec_multiple}.

A generating functional for correlation functions of densities and currents in the thermal state $\rho_\beta$ can be written
\begin{equation}\label{eq_Z}
Z[A^1,A^2] = {\rm Tr}\left( U[A^1]\rho_\beta U^\dagger[A^2]\right), 
\end{equation}
where $A^1,\,A^2$ are background gauge fields that couple to the conserved current in the time-evolution operator
\begin{equation}
U[A]
	= \mathcal T \exp \left\{-i \int_{-\infty}^\infty\!\! dt \left(H - \int d^dx\, j^\mu A_\mu(t,x)\right)\right\}\, , 
\end{equation}
where we have collectively denoted the charge and current density by $j^\mu = (n,j^i)$. Derivatives of $\log Z$ with respect to $A^1,\, A^2$ can generate correlation functions of $j^\mu$ with various time orderings. If the system thermalizes, one expects the partition function to have a representation in terms of a local effective Lagrangian of the long-lived hydrodynamic variables. It is local in space and time because there are no other long-lived excitations in the thermal state---this is the assumption of thermalization. In the approach of Ref.~\cite{Crossley:2015evo}, this effective Lagrangian is a function of the fluctuating density $n$, and a conjugate field $\phi_a$:
\begin{equation}\label{eq_Z_EFT}
Z[A^1,A^2]
	\simeq \int Dn D\phi_a \, e^{i \int dt d^d x \mathcal L}\, .
\end{equation}
What is gained in universality is lost in exactness: while it is not an exact representation of the microscopic partition function \eqref{eq_Z}, Eq.~\eqref{eq_Z_EFT} provides a systematic expansion for it when background fields $A$ have slow variation in time (and space) compared to the local equilibration time of the system. We further motivate this construction in Appendix~\ref{app_EFTdetails}, and focus here on how it is used to obtain universal corrections to diffusion. To leading order in derivatives, the effective Lagrangian is found to be
\begin{equation}\label{eq_L_EFT}
\mathcal L
	= i \sigma(n) \left(\nabla \phi_a\right)^2 - \phi_a \left(\dot n - \nabla (D(n)\nabla n)\vphantom{n^2_a}\right) + \cdots
\end{equation}
Here, $\sigma(n)$ and $D(n)$ are functions of the density that are not fixed by the EFT: they correspond to the conductivity and diffusivity of the system. These functions also play an important role in macroscopic fluctuation theory \cite{RevModPhys.87.593}. In the present approach, they are just the leading terms in a general expansion in derivatives (see for example \cite{Jain:2020zhu} for a discussion of certain terms in the EFT that do not appear in constitutive relations).

As in most EFTs, it is typically impossible to derive \eqref{eq_Z_EFT} from a microscopic model of interest. One exception is in the context of strongly interacting holographic quantum field theories, where progress has been made in deriving at least the quadratic part of the EFT from microscopics \cite{deBoer:2018qqm,Glorioso:2018mmw,Ghosh:2020lel} (see \cite{Policastro:2002se,Herzog:2002pc,Skenderis:2008dg,Nickel:2010pr} for earlier work in this direction); similar derivations may be possible for lattice systems with large local Hilbert space dimension (e.g., \cite{Khemani:2017nda,Chan:2017kzq}), or noisy systems in the limit of strong noise (e.g., \cite{bauer2017stochastic,Claeys:2021skz}).

When studying linear response or more general correlation functions, one expands these functions around the background value of interest for the density $n = \bar n + \delta n$
\begin{equation}\label{eq_D_exp}
D(n)
	\simeq D + \delta n D'  + \frac12 \delta n^2 D'' + \cdots\, , 
\end{equation}
where $D,D',D'',$ etc.~on the right-hand side are evaluated at the background density $\bar n$. These are {\it Wilsonian coefficients} of the EFT: the are not fixed by the EFT (and in fact are not universal), but the EFT instead predicts how they enter in any late time observable. Since the same coefficients enter in a large number of observables, the problem is highly overdetermined and the EFT has substantial predictive power.

In the following subsections, we use the EFT \eqref{eq_L_EFT} to compute 1-loop and 2-loop corrections to the retarded Green's function of the charge density
\begin{equation}\label{eq_GR}
G^{R}(\omega,q)
	= \frac{\sigma q^2}{-i\omega + D q^2} + \delta G^R(\omega,q)\, .
\end{equation}
$G^{R}$ is simply related to the Fourier transform of the dynamical structure factor through fluctuation-dissipation relations, but has a simpler analytic structure and is therefore more convenient to work with. In the EFT, it can be obtained from the mixed correlator (see App.~\ref{app_EFTdetails})
\begin{equation}\label{eq_nphia_to_GR}
G^R(\omega,q)
	= i \sigma q^2 \langle n \phi_a\rangle(\omega,q)\, .
\end{equation}
At tree-level, this can be evaluated using the propagators of the fields obtained from the Gaussian Lagrangian \eqref{eq_L_EFT}:
\begin{equation}\label{eq_propagators}
\begin{split}
\langle n \phi_a\rangle_0(\omega,q) 
	&= \frac{1}{\omega+ i D q^2}\, ,\\
\langle n n\rangle_0(\omega,q)
	&= \frac{2\sigma q^2}{\omega^2 + D^2 q^4}\, .
\end{split}
\end{equation}
Using \eqref{eq_nphia_to_GR}, one recovers the leading diffusive behavior in \eqref{eq_GR}. The second piece $\delta G^R(\omega,q)$ comes from loop and higher derivative corrections, which are studied below.

\subsection{1-loop}

Loop corrections to \eqref{eq_GR} arise due to nonlinearities in the EFT. For example, expanding $D(n)$ as in \eqref{eq_D_exp} leads to a cubic term 
\begin{equation}\label{eq_L3}
\mathcal L^{(3)}
	= \frac12 D' \nabla^2 \phi_a n^2\, .
\end{equation}
This produces a cubic vertex which, working perturbatively in these interactions, will lead to loop corrections to $G^R$. Note that the perturbative expansion is always controlled, because nonlinearities are irrelevant. Indeed, \eqref{eq_propagators} implies that density fluctuations scale as $\delta n(t,x) \sim q^{d/2}$; since the cubic nonliniearity is suppressed by an extra power of $\delta n$, it gives small corrections at late times or long distances, where $\omega\sim q^2 \to 0$. This is in contrast to momentum conserving systems in $d=1$, where nonlinearities are relevant and lead to a breakdown of diffusion which is replaced by the KPZ universality class \cite{PhysRevA.16.732}. That the perturbative expansion is controlled in the present situation is a derivation of the EFT, rather than an assumption. 

The cubic action also contains a term proportional to $\sigma'$. While this term leads to nonlinear response \cite{Delacretaz:2023ypv}, in Appendix~\ref{ap:A3} we show that it does not contribute to the 1-loop corrected two-point function; we therefore ignore it here.

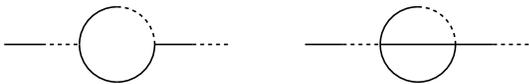
\begin{figure} 
	\begin{subfigure}{0.45\columnwidth}
		\centering
		\scalebox{0.5}{
		\begin{tikzpicture}
		\begin{feynman} \vertex at (0,0) (i1);
		\vertex at (1,0) (i2);
		\vertex at (2,0) (i3);
		\vertex at (3,1) (i4);
		\vertex at (3,-1) (I4);
		\vertex at (4,0) (i5);
		\vertex at (5,0) (i6);
		\vertex at (6,0) (i7);
		\diagram*{
			(i1) -- [plain, very thick] (i2),
			(i2) -- [scalar, very thick] (i3),
			(i3) -- [plain,quarter left, very thick] (i4),
			(i4) -- [scalar,quarter left, very thick] (i5),
			(i3) -- [plain,quarter right, very thick] (I4),
			(I4) -- [plain,quarter right, very thick] (i5),
			(i5) -- [plain, very thick] (i6),
			(i6) -- [scalar, very thick] (i7)
		};
		\end{feynman}
		\end{tikzpicture}
		}
	\end{subfigure}
	\begin{subfigure}{.45\columnwidth}
		\centering
		\scalebox{0.5}{
		\begin{tikzpicture}
		\begin{feynman} \vertex at (0,0) (i1);
		\vertex at (1,0) (i2);
		\vertex at (2,0) (i3);
		\vertex at (3,1) (i4);
		\vertex at (3,-1) (I4);
		\vertex at (4,0) (i5);
		\vertex at (5,0) (i6);
		\vertex at (6,0) (i7);
		\diagram*{
			(i1) -- [plain, very thick] (i2),
			(i2) -- [scalar, very thick] (i3),
			(i3) -- [plain,quarter left, very thick] (i4),
			(i4) -- [scalar,quarter left, very thick] (i5),
			(i3) -- [plain,quarter right, very thick] (I4),
			(I4) -- [plain,quarter right, very thick] (i5),
			(i3) -- [plain, very thick] (i5),
			(i5) -- [plain, very thick] (i6),
			(i6) -- [scalar, very thick] (i7)
		};
		\end{feynman}
		\end{tikzpicture}
		}
	\end{subfigure}
\caption{\label{fig_12loop}{\em Left:} 1-loop correction to diffusion. {\em Right:} 2-loop correction to diffusion at half-filling. The propagators $\langle nn\rangle_0(\omega,q)$ and $\langle n\phi_a\rangle_0(\omega,q)$ from Eq.~\eqref{eq_propagators} correspond to the solid lines, and half-solid half-dashed lines respectively.}
\end{figure}

The cubic vertex \eqref{eq_L3} leads to a 1-loop correction to $\langle n\phi_a\rangle$ shown in Fig.~\ref{fig_12loop}. Its evaluation, performed in Ref.~\cite{Chen-Lin:2018kfl}, is streamlined here. It is convenient to amputate the external legs and parametrize the correction as $\delta D(\omega,q)$, namely:
\begin{equation}
\delta \langle n \phi_a\rangle
	= -i q^2 \delta D(\omega,q) \bigl(\langle n\phi_a\rangle_0 \bigr)^2 \, .
\end{equation}
The one-loop correction then takes the form
\begin{equation}\label{eq_deltaD_1loop}
\delta D(\omega,q)
	= -i D'^2 \int_{p'} q'^2 \langle n\phi_a\rangle(p') \langle nn\rangle(p-p')
\end{equation}
where we have used the short-hand notation $p\equiv\{\omega,q\}$, and $\int_p \equiv \int \frac{d\omega d^d q}{(2\pi)^{d+1}}$. The loop integrals can be readily evaluated in any dimension (see Appendix~\ref{ap:A3}), and give
\begin{equation}\label{eq_deltaD_1loop_result}
\delta D(\omega,q)
	= \frac{\chi D'^2}{D^2} (-i\omega) \alpha_d\left(q^2 - \frac{2i\omega}{D}\right)\, , 
\end{equation}
where $\chi\equiv \sigma/D$ is the static susceptibility, and with 
\begin{equation}\label{eq_alphad}
\alpha_d(z)
	= \frac{(-z)^{\frac{d}2-1}}{(16\pi)^{d/2}\Gamma (\frac{d}{2})}  \cdot
\begin{cases}
i\pi	& \hbox{if $d$ odd}\, ,\\
\log\frac1{z}  		& \hbox{if $d$ even} \, .
\end{cases}
\end{equation}
The general scaling $\delta D/D \sim q^{d}$ agrees with expectations: the cubic interaction is suppressed by $\delta n \sim q^{d/2}$, and two cubic vertices were necessary to produce a loop correction. The detailed loop calculation was necessary to obtain the overall coefficient, as well as the entire dependence on the dimensionless ratio $D q^2/\omega$. Nevertheless, several aspects of the result could have been anticipated on general grounds: (i) the fact that the correction vanishes in the static limit $\lim_{\omega\to 0} \delta D(\omega,q)=0$ is required by the analyticity of equilibrium thermal correlators due to the finite thermal correlation length \cite{Jain:2020hcu}, 
(ii) the existence of a branch point at $\omega = -\frac{i}{2} D q^2$ follows from a simple cutting argument \cite{Chen-Lin:2018kfl}.

We are most interested in the case $d=1$:
\begin{equation}\label{eq_deltaD_1loop_d1}
\delta D(\omega,q)
	= \frac{\chi D'^2}{D^2} (-i\omega)\frac{1}{4} \frac{1}{\sqrt{q^2 - \frac{2i\omega}{D}}}\, .
\end{equation}
Fourier transforming this expression, or rather $\delta G^R(\omega,q)$, is straightforward, but presented in Appendix~\ref{ap:FT} for completeness; this results in a correction to the correlation function shown in Eq.~\eqref{eq_F10}.

\subsection{2-loop--half filling}

When $D'=0$, there is no one-loop correction to diffusion. 
This situation naturally arises in particle-hole symmetric systems at half filling, because $D'\equiv \frac{dD(n)}{d\delta n}$ is odd under particle-hole (or charge conjugation) symmetry $\delta n\to -\delta n$ and must therefore vanish.
The leading fluctuation corrections come instead from a two-loop diagram, shown in Fig.~\ref{fig_12loop} arising from the quartic vertex
\begin{equation}
\mathcal L^{(4)}
	= \frac{1}{6} D'' \nabla^2 \phi_a n^3\, .
\end{equation}
Because this interaction scales as $\mathcal L^{(4)}/\mathcal L^{(2)}\sim \delta n^2 \sim q^d$, and that two such vertices will be necessary to give a non-analytic correction to the two-point function, the two-loop correction will scale as $\delta G^R/G^R \sim q^{2d} \sim \omega^d$ (up to logarithms). In $d=1$, these corrections are as large as higher derivative corrections to diffusion, studied in the next section, which scale as $q^2$.

The two-loop correction to $\langle n \phi_a\rangle$, with external legs amputated, is given by
\begin{align}\label{eq_deltaD_2loop}
&\delta D(\omega,q)
	= - \frac{i}{2}D''^2 \\ \notag
&\quad \times\int_{p',p''}q''^2 \langle n \phi_a\rangle(p'') \langle nn\rangle(p'-p'') \langle nn\rangle(p-p')\, .
\end{align}
The integrals are evaluated in Appendix~\ref{ap:2loop}. One finds
\begin{equation}\label{eq_deltaD_2loop_result}
\delta D(\omega,q)
	= \frac{1}{2} \frac{(\chi D'')^2}{D^2} (-i\omega) \beta_d \left(q^2 - \frac{3i\omega}{D}\right)\, , 
\end{equation}
with
\begin{equation}
\beta_d(z)
	= \frac{(-z)^{d-1} \log \frac1z}{(12\sqrt{3}\pi)^d \Gamma(d)}\, .
\end{equation}
This result has the expected $q^{2d}$ scaling, vanishes in the static limit $\lim_{\omega\to 0} \delta D(\omega,q)=0$, and features the expected 3-diffuson branch point at $\omega = - \frac{i}{3} D q^2$ \cite{Delacretaz:2020nit}.

For $d=1$, this becomes
\begin{equation}\label{eq_deltaD_2loop_d1}
\delta D(\omega,q)
	= \frac{1}{24 \sqrt{3} \pi} \frac{(\chi D'')^2}{D^2}(-i\omega) \log \frac{1}{q^2 - \frac{3i\omega}{D}}\, , 
\end{equation}
leading to a correction to $G^R(\omega,q)$ whose Fourier transform is computed in \eqref{eq_FT_tmp} and shown in \eqref{eq_F20}.

\subsection{Higher-derivative corrections}\label{ssec_higherderiv}

Higher derivative corrections are also captured by the EFT for diffusion \eqref{eq_L_EFT}. These will either involve extra time derivatives $\partial_t$, or two extra spatial derivatives $\nabla^2$ (by reflection symmetry), and therefore give corrections to the leading behavior that are suppressed by $q^2$, or equivalently $1/t$. 
One can write the most general such higher derivative corrections to the EFT (see \cite{Crossley:2015evo}). However, since we are interested in the two-point function, we can instead directly write the most general corrections to $G^R$. The higher derivative corrections should be treated perturbatively, as quadratic vertices; the final expression therefore contains at most two powers of the diffusive propagator. The most general $O(q^2)$ correction to the retarded Green's function is therefore
\begin{equation}\label{eq_GR_q2}
\begin{split}
G^R(\omega,q)
	&= \frac{\sigma q^2}{-i\omega + Dq^2} \left[1 + c_1 (-i\omega) + c_2\frac{(-i\omega)^2}{-i\omega + D q^2}\right]\\ 
	& + \tilde c_1 q^2  + \tilde c_2 (-i\omega) + O (q^4) \, .\\[-1em]
\end{split}
\end{equation}
The two coefficients $\tilde c_1,\,\tilde c_2$ are contact terms and will not affect the correlation function at separated points: $\tilde c_1$ has the interpretation of a $q^2$ correction to the static susceptibility $\chi(q) = \chi + \tilde c_1 q^2 + \cdots$, and $\tilde c_2$ is in fact forced to vanish to guarantee $G^R(\omega,q\to 0)=0$. Fourier transforming $\langle nn\rangle = \frac{2}{1-e^{-\beta\omega}}\Im G^R(\omega,q)$ leads to \eqref{eq_F01} (we have taken the liberty to redefine the non-universal coefficients $c_1,c_2$). 

Eq.~\eqref{eq_GR_q2} can also be derived through more conventional approaches to hydrodynamics \cite{kadanoff1963hydrodynamic,chaikin1995principles}: one writes the linearized constitutive relation for the current in terms of the charge $\delta n$ up to subleading order in derivatives
\begin{equation}
j_i(t,x) = 
    -D\partial_i \delta n + D_2 \nabla^2 \partial_i \delta n + \cdots\, .
\end{equation}
We have omitted a term $\partial_t\partial_i \delta n$ which would have the same scaling as $D_2$, because it can be absorbed in $D_2$ using the leading equations of motion $\partial_t\delta n = D \nabla^2 \delta n$.
To obtain response functions, one needs to know the constitutive relation in the presence of a source $\delta \mu(t,x)$ for charge density. Assuming that the equilibrium response is given by $\delta n(q) = \chi(q) \delta \mu(q)$, with $\chi(q)\simeq \chi + \chi_2 q^2 + \cdots$ the static susceptibility, the current constitutive relation in the presence of sources must take the form (in momentum space)
\begin{equation}
\begin{split}
j_i(t,q) = 
    &-iq_i(D + D_2 q^2) (\delta n - (\chi+\chi_2 q^2)\delta \mu)\\
    &+ \gamma iq_i \partial_t \delta \mu + \cdots\, ,
\end{split}
\end{equation}
where the combination $\delta n - \chi(q)\delta \mu$ in the first line is required for the current to vanish in thermal equilibrium. Note however that this argument allows for terms involving the time derivative of the source, as in the second line. Inserting the current in the continuity relation $\partial_t n + \partial_i j_i = 0$ and solving for $\delta n$ yields a retarded Green's function $G^R(\omega,q) \equiv \frac{\delta n(\omega,q)}{\delta \mu(\omega,q)}$ that matches \eqref{eq_GR_q2} with $\chi_2 = \tilde c_1,\, D_2 = -D^2 c_2, \, \gamma=\tilde c_1 + (c_1+c_2)\sigma$.

\subsection{Multiple densities}\label{sec_multiple}

Systems with multiple conserved densities can be studied similarly, by including all densities in the EFT. The general scaling of loop corrections remains unchanged; however, mixing of the densities allows for new scaling functions with qualitatively different features. Indeed, consider for example systems with conservation of both charge $\dot n + \nabla\cdot j = 0$ and energy (or heat) $\dot \varepsilon + \nabla\cdot j^{\varepsilon} = 0$. Nonlinearities can now involve both densities, e.g.:
\begin{equation}\label{eq_j_heatcharge}
j_i = \cdots + \lambda \delta \varepsilon \partial_i  \delta n + \cdots\, .
\end{equation}
The coefficient $\lambda$ arises from a temperature-dependent conductivity $\partial_T\sigma$ (or, equivalently, a density-dependent thermoelectric conductivity $\partial_\mu \alpha$). While it seems similar to the single density nonlinearity $D'$ considered above, this term is qualitatively different because it is not a total derivative contribution to the current. It therefore contributions to the $q=0$ optical conductivity $\sigma(\omega)\sim 1 + \lambda^2|\omega|^{d/2} + \cdots$, as was already recognized in Ref.~\cite{PhysRevB.73.035113}.

In order to obtain the universal scaling functions at finite $q$, necessary to make predictions for the structure function $\langle n(t,x)n\rangle$, the EFT is generalized to systems with multiple conservation laws in App.~\ref{app_mult}. The scaling functions are in this case complicated by the fact that there are several diffusivities, and therefore several natural scaling variables $y=x/\sqrt{Dt}$. To illustrate the appearance of novel scaling functions with multiple densities in a simple context, we focus on the hydrodynamics of densities for a non-abelian internal symmetry, say $SU(2)$. This situation is simpler because the $SU(2)$ symmetry restricts the susceptibilities to be diagonal $\chi_{AB} \equiv \frac{dn_A}{d\mu^B} = \chi \delta_{AB}$, leads to a single diffusivity $D$, and only allows for one cubic nonlinearity in the EFT which has a clear similarity with Eq.~\eqref{eq_j_heatcharge}:
\begin{equation}
j_i^A = - D \partial_i n^A + \lambda\epsilon_{ABC} n^B \nabla n^C + \cdots\, .
\end{equation}
Here $A,B,C,\ldots$ run over the three elements of the $SU(2)$ algebra.
The hydrodynamic description of thermalizing systems with non-abelian internal symmetries has been studied before \cite{Son:2002ci,Torabian:2009qk,Fernandez-Melgarejo:2016xiv,Glorioso:2020loc,Claeys:2021skz}, with the role of the nonlinearity $\lambda$ particularly emphasized in \cite{Glorioso:2020loc,Claeys:2021skz}. In Appendix~\ref{app_mult}, we show that the 1-loop correction to the density two-point function is, in one spatial dimension
\begin{align}
&G^R_{n^An^B}(\omega,q) = \delta_{AB} G^R(\omega,q)\, , \\ \notag
&G^R(\omega,q) =\frac{\sigma q^2}{ D q^2 -i\omega }
	+ \frac{\lambda^2 \chi^2}{D} i\omega q^2 \frac{\sqrt{q^2 - \frac{2i\omega}{D}}}{\left(Dq^2 - i\omega\right)^2} + \cdots \, .
\end{align}
This produces a non-analytic correction at small frequencies to the optical conductivity
\begin{equation}
\begin{split}
\sigma(\omega)
	&= \lim_{q\to 0} \frac{-i\omega}{q^2} G^R(\omega,q)\\
	&= \sigma - \frac{\lambda^2 \chi^2}{D^{3/2}} (1-i)\sqrt{\omega} + \cdots\, .
\end{split}
\end{equation}

The correction to the density two-point function in spacetime domain can be found by Fourier transforming (see Sec.~\ref{app_mult}). One finds a correction similar to \eqref{eq_nn_noph}, with a different universal scaling function
\begin{align}\label{eq_nn_su2}
F_{1,0}^{\rm mult}(y)
	&= \frac{\chi \lambda^2}{D^{5/2}} \tilde F_{1,0}^{\rm mult}(y) \\ \notag
\tilde F_{1,0}^{\rm mult}(y) 
	&= \frac{4-y^2}{4\sqrt{\pi}} e^{-y^2/2} + \frac{y(2-y^2)}{8} e^{-y^2/4} {\rm Erf}(y/2)\,.
\end{align}
%


\subsection{Confirming the EFT with classical numerics}\label{ssec_KLS}

Before turning to quantum simulations, where the limited accessible time scales make it crucial to account for power-law corrections to diffusion, we confirm the EFT predictions in classical thermalizing systems. We focus on classical lattice gases satisfying the `gradient condition', namely where the current density is a total derivative microscopically. In these situations, the diffusivity $D(n)$ is known analytically \cite{spohn2012large}, making it simple to perform precision tests of EFT predictions  \cite{Delacretaz:2023ypv}. Indeed, since the loop corrections \eqref{eq_nn_noph} and \eqref{eq_nn_ph} only depend on the susceptibility $\chi$ and derivatives of $D(\rho)$, they are entirely fixed analytically and can be directly compared to numerics.

\begin{figure}
	\centering
\includegraphics[width=0.95\columnwidth]{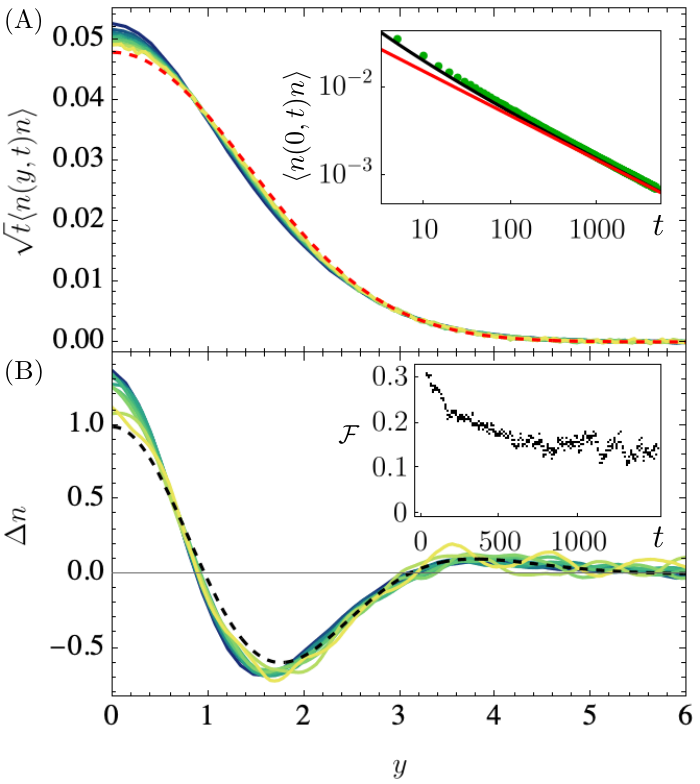}\caption{\label{Fig:3} (A): Profile of the dynamical structure factor for the KLS model with parameters $\delta = \rho = 0.9, \epsilon = 0$. Different coloured curves denote different times $t \in \{200,2000\}$ with smaller times corresponding to darker colours. The red dashed curve is the diffusive prediction $\frac{\chi}{\sqrt{4\pi D}}F_{0,0}(y)$. Inset: Autocorrelation function ($y=0$). Diffusive predictions with (black), and without (red), leading order corrections, Eq.~(\ref{eq:autocor}) are  shown. (B): Comparison between the correction to diffusion from simulation data $\Delta n \equiv \sqrt{t}\left[\langle n(x,t)n\rangle - \frac{\chi}{\sqrt{4\pi D t}} F_{0,0}(y)\right]$ and the EFT prediction $F_{1,0}(y)$ (black, dashed), Eq.~(\ref{eq_nn_noph}). Inset: Absolute area between the finite time curves and the analytic expression, $\mathcal{F}= \int^3_0 dy|\Delta n (t)- F_{1,0}|/\int^3_0 dy |F_{1,0}|$.}
\end{figure}

As a simple example of a lattice gas satisfying the gradient condition with a non-trivial $D(\rho)$, we consider the one-dimensional Katz-Lebowitz-Spohn model \cite{katz1984nonequilibrium,PhysRevE.63.056110,PhysRevLett.118.030604}, describing a collection of hard core particles hopping on a lattice with rates depending on the occupation of neighbors:
\begin{subequations}
\begin{align}
0\, 1\, 0\, 0 \quad \xrightarrow{r(1+\delta)} \quad 0\, 0\, 1\, 0 \\
1\, 1\, 0\, 1 \quad \xrightarrow{r(1-\delta)} \quad 1\, 0\, 1\, 1 \\
1\, 1\, 0\, 0 \quad \xrightarrow{r(1+\epsilon)} \quad 1\, 0\, 1\, 0 \\
0\, 1\, 0\, 1 \quad \xrightarrow{r(1-\epsilon)} \quad 0\, 0\, 1\, 1 
\end{align}
\end{subequations}
with equal rates for the spatially reversed processes. $\delta$ and $\epsilon$ are two parameters of the model, whereas $r$ defines the unit for time and can be set to unity. We focus on the model with $\epsilon=0$, corresponding to infinite temperature dynamics, which allows to use a random initial state as a thermal state (taking $\epsilon\neq 0$ instead requires prethermalizing the system, making numerics more costly). In this situation, the susceptibility is $\chi(\rho) = \rho(1-\rho)$ and diffusivity $D(n) = 1+\delta(1-2\rho)$, so that $D' = -2\delta$. This fixes all parameters entering in the leading correction to diffusion, Eqs.~\eqref{eq_nn_noph}, \eqref{eq_F10}. Fig.~\ref{Fig:3} shows the excellent agreement between the EFT prediction and numerics. We stress that the entire scaling {\em function} agrees quantitatively with the 1-loop prediction $F_{1,0}(y)$, and that no fitting parameter is involved in this comparison. 

\begin{figure*}[t]
\begin{center}
	\includegraphics[width=2.\columnwidth]{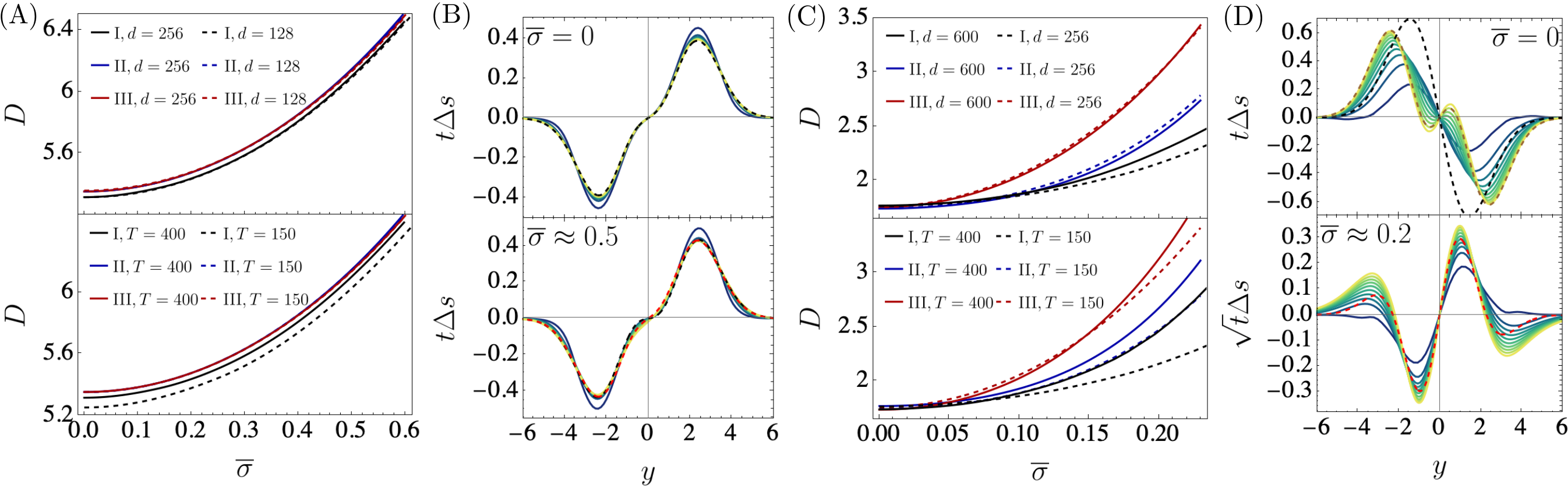}
  \caption{\label{Fig:4} Tensor network simulations of driven XXZ chain with (A,B) decoherence, $\gamma = 0.1$, and (C,D) staggered field, $g = 0.4$. (A,C) Diffusivity as a function of equilibrium magnetization. Top: different bond dimensions, $d$. Bottom: different simulated times $T$, using the same bond dimension $d = 256$. Different colours denote fitting methods which take into account an increasing amount of corrections to leading order  diffusion, I$\rightarrow$ II$\rightarrow$III, see Appendix~(\ref{sec:fit}).  (B,D) Corrections to diffusion for $d = 256$, evaluated at different times $t = (40,80,\ldots,400)$, denoted by dark blue$\rightarrow$yellow colors.  Black dashed line denotes linear corrections \textbf{F}$_{0,1}$. Red dashed line denotes \textbf{F}$_{1,0}$ in (D), and  \textbf{F}$_{0,1}+\sqrt{T}$\textbf{F}$_{1,0}$ with $T= 400$ in (B). The brown dashed line in (D) denotes the combined effect of 2-loop correction and linear corrections \textbf{F}$_{0,2}+$\textbf{F}$_{1,0}$, where the logarithmic in time component of \textbf{F}$_{0,2}$ is evaluated at $T= 400$.}
  \end{center}
\end{figure*}

Analytical knowledge of the leading order correction allows for considerably improved predictions for diffusivity when the available integration time is short, i.e. when 1-loop effects are strong. Both experiments and simulations, especially in quantum systems, are usually limited to relatively short timescales, and uncontrolled extrapolations are therefore employed to get infinite time properties such as diffusivity. We propose a robust method which takes into account 1-loop effects by fitting the three-dimensional dataset $\langle n(x,t)n\rangle_\rho$ versus $t,x,\rho$ with Eq.~(\ref{eq_nn_noph}). We compare this method to the fit using just the leading order term (equivalently taking $F_{1,0} = 0$). For example, we aim to approximate $D(\delta = 0.9)= 1.9 - 1.8 \rho$ around $\rho = 0.9$. For that, we simulate a sample of densities $\rho = (0.85,0.86,\ldots 0.93)$. The diffusivity is parametrized by $D_{\rm fit}= a - b \rho$ since we know its analytic form. In general, the parametrization may include additional powers of density as the precise form of diffusivity is not polynomial. We constrain the time window  $t= 40-T$, $T= 100$ and find the deviations of the fit from the exact diffusivity: $\frac{1.9- a}{1.9} \approx (0.135, 0.081)$ and $\frac{1.8- b}{1.8} \approx (0.12, 0.084)$, where the first/second number in the parenthesis denote the fit without/with the 1-loop correction. Our results show a quantitative improvement which increases as $T$ decreases and 1-loop effects become stronger. In principle, it is possible to perform time extrapolations to the above fitting method by using time windows of varying size.

\section{Quantum Transport}\label{sec_QT}
To test and make use of the EFT predictions in a minimal setting, we focus on quantum-coherent and incoherent chaotic systems with a single conserved charge. We estimate the dependence of diffusivity on the equilibrium magnetization using various approaches, and show that the corrections to diffusion are in agreement with EFT. Moreover,  incorporating them in the fitting methods can significantly improve the  diffusivity approximation at finite times. 

\subsection{Model and Methods}

The conserved charge is chosen to be magnetization,
\begin{equation}
    N = \sum^L_{i=1}  \sigma^z_i,
\end{equation}
where $\sigma^z$ is the Pauli-z matrix and $L$ is the system size. Since magnetization is a sum of local operators, the equilibrium ensemble is a product state, i.e., $e^{\mu N} = \prod^L_{i=1}e^{\mu \sigma^z_i}$. We did not consider systems with charges such as energy density which have equilibrium states with a finite correlation length because this adds an additional layer of complexity to the simulations, even though the EFT predictions are the same.

As a minimal chaotic model where magnetization is the only conserved quantity, we choose the Floquet-XXZ chain with a staggered field, whose stroboscopic dynamics are generated by a Floquet operator
\begin{align}\label{eq:unitary}
&U = U_{e} U_{o}, \quad U_{e} = \prod^{L/2}_{i=1}U_{2i,2i-1},\;U_{o} = \prod^{L/2}_{i=1}U_{2i-1,2i-2}\notag\\
&U_{i,i+1}= \text{exp}\left(-i (h^{XXZ}_{i,i+1} + V_{i,i+1})\right). 
\end{align}
The evolution is performed by first evolving the odd bonds and then the even bonds, with two-body gates generated by the following operators,
\begin{equation}
    \begin{split}
        &h^{XXZ}_{i,i+1} = J (\sigma^+_i\sigma^-_{i+1} + {\rm h.c.}) + \frac{\Delta}{2}\sigma^z_i\sigma^z_{i+1},\\
        &V_{i, i+1} = g \left((-1)^i \sigma^z_i + (-1)^{i+1} \sigma^z_{i+1}\right).
    \end{split}
\end{equation}
In the absence of a staggered field $V$ the Floquet-XXZ chain is integrable~\cite{Gritsev17integrable}. For our choice of parameters, $J = \pi/4, \Delta = J - 0.2$, the magnetization at $V = 0$ displays ballistic transport. Turning on staggered field leads to integrability breaking, the system becomes chaotic, and therefore magnetization is expected to diffuse. To establish our method we alternatively perturb the XXZ-chain with Markovian noise (dephasing). Dephasing effectively suppresses the generation of operator entanglement in the simulation, leading to very accurate numerical data. To simulate the  noise-averaged state, we define the dephasing map by the action of the  local channel on the state of a single spin,
\begin{equation}
    \qquad \mathcal{D}_i\left(\begin{array}{cc}
\rho_{1,1} & \rho_{1,0}\\
  \rho_{0,1}&  \rho_{0,0}
\end{array}\right)= \left(\begin{array}{cc}
\rho_{1,1} & e^{-\gamma }\rho_{1,0}\\
 e^{-\gamma }\rho_{0,1}&   \rho_{0,0}
 \end{array}\right).
\end{equation}
The global noise channel is a product of local channels and is applied to the state following a period of coherent evolution,
\begin{equation}\label{eq:decoherence}
\mathcal{D} = \bigotimes^{N}_{i=1}\mathcal{D}_i, \quad \rho(t+1) = \mathcal{D}(U\rho (t)U^{\dag}).
\end{equation}

To study the linear response dynamics, we employ the weak domain wall initial state proposed by Ljubotina et al.~\cite{Prozen19PRL},
\begin{equation}\label{eq:dw}
\rho(\mu ,\delta , t=0) =\frac{1}{M} e^{\mu N}\left(\prod^{L/2}_{i=1}e^{-\delta \sigma^z_i} \bigotimes \prod^{L}_{i=\frac{L}{2}+1}e^{\delta \sigma^z_i}\right),
\end{equation}
where $M=\text{tr} \rho $ is the normalization constant and $\delta \rightarrow 0$ generates a weak domain wall perturbation on top of the equilibrium state characterized by the chemical potential $\mu$. The linear response regime is characterized by a  quench where the amount of injected magnetization is not extensive. Inspired by the leading non-linear correction, Eq.~\ref{eq_nn_noph}, a natural condition for the linear response regime is $\delta \ll \frac{\chi D'}{\sqrt{t_{\rm max}}D}$, where $t_{\rm max}$ is the maximum simulation time. We use $\delta = 0.0005$, which satisfies the condition and is also numerically checked to be in the linear response limit for all simulated times.

The Floquet evolution defined by Eq~(\ref{eq:unitary}) breaks translation invariance, since even and odd sites are not equivalent. To simplify the analysis, we average over even and odd sites. In addition, we shift the magnetization by its equilibrium value, $\overline{\sigma} = \text{tr} \left(\sigma^z_i\rho(\mu,0)\right)=\tanh {\mu}$ and normalize its initial magnitude to $1/2$,
\begin{equation}\label{eq:mag_resc}
    s_{j} = \frac{\text{tr} \left( \sigma^z_{2 j-1} \rho \right)+\text{tr} \left( \sigma^z_{2j} \rho \right) - 2\overline{\sigma}}{4  | \text{tr} \left( \sigma^z_{1} \rho(t=0) \right) |},
\end{equation}
where $j\in \{1 ,2,\ldots,\overline{L}\}$, and $\overline{L}=L/2$. In this  normalization, the initial state profile is $ s_{j\leq\overline{L}/2} = -0.5$ and $  s_{j>\overline{L}/2} = 0.5$ for all values of  $\mu$ and $\delta$. Since we doubled the lattice spacing, diffusivity and static susceptibility, $\chi = \frac{1}{L}\left(\langle N^2 \rangle - \langle N \rangle^2 \right) = (\cosh\mu)^{-2}$, are rescaled accordingly, $D \rightarrow D/4$, $\chi \rightarrow \chi/2$. We will always present the results of diffusivity computed using the original lattice spacing.  For clarity, we employ a continuum description of the lattice variables $s_j \rightarrow s(x)$, since the hydrodynamic corrections are defined in continuum.

A simple relation between the domain-wall quench for $\delta \rightarrow 0$ and the dynamical structure factor was originally derived in Ref.~\cite{ljubotina2017spin}. In the continuum limit, and under the conventions described in the previous paragraph, the relation simply reads $\langle \sigma_z(x,t)\sigma_z\rangle_c = \chi \frac{d s}{dx}$, where the subscript $c$ stands for the connected part of the correlation function and the average is performed over the equilibrium state, $\rho(\mu ,\delta = 0)$. Distance $x$ is measured from the position of the domain wall at $\overline{L}/2$. While the two quenches are formally equivalent, we prefer to transform (simple integral in space) the EFT results to the domain-wall picture because taking the spatial derivative (or directly calculating the dynamical structure factor) of the numerical data enhances the  errors generated by the simulations. Therefore, the magnetization profile in the domain-wall quench takes form,
\begin{equation}\label{eq:EFT_DW}
\begin{split}
 &\mathcal{S}(x,t) =  \sum_{m,n}\frac{1}{t^{m/2+n}}\mathbf{F}_{m,n},\\
   & \mathbf{F}_{m,n} = \frac{1}{\sqrt{4\pi D t}}\int_{0}^x dx F_{m,n},
    \end{split}
\end{equation}
where $F_{m,n}$ are the functions described in Section~\ref{sec_scalingfunctions}. The leading order diffusion is given by $\mathbf{F}_{0,0} = \frac{1}{2}\text{Erf}(y/2)$ and the corrections are presented in Eq.~(\ref{eq:EFT_eqs_dw}). 

The diffusive corrections are numerically explored by simulating the dynamics of large system sizes, using tensor network techniques. We employ the matrix product density operator (MPDO)~\cite{VerstraetePRL04,Vidal2004PRL} representation of the state, and evolve it with time evolving block decimation (TEBD)~\cite{VidalPRL03} algorithm. The simulations are performed using the ITensor library~\cite{itensor}. We simulate the dynamics from equlibria states with chemical potentials $\mu = (0,0.01,\ldots)$. In our simulations we evolve up to a time $T = 400$ and fix the system size to $L = 2 T$ in order to avoid finite size effects. The numerical results are shown to be convergent for the bond dimensions employed.

To probe the strength of the non-linear corrections we calculate the prefactor of \textbf{F$_{1,0}$} defined in Eq.~(\ref{eq:EFT_eqs_dw}),
\begin{equation}
   C_1=  \frac{\chi D'^2}{\sqrt{4 \pi D^5}},
\end{equation}
where $D' = \frac{d D}{d\overline{\sigma}}$ and the magnetization profile after subtracting the leading order term,
\begin{equation}
    \Delta s(y) = s(y) - \textbf{F$_{0,0}$}.
\end{equation}

The estimation of diffusivity $D(\overline{\sigma})$, is achieved by three different fitting schemes, labelled (I, II, III). In the following, we present a brief overview of each scheme and further elaborate on the novel method II, which for reasons that will become clear later, is more accurate than leading order diffusion fitting and can be applied to any diffusive system efficiently. Methods I and III are elaborated in  Appendix~\ref{Ap:QC}. I is a scheme in which the approximate diffusivity is extracted by fitting the dynamical structure factor at the largest time with the leading order profile $\textbf{F}_{0,0}$. II, additionally  takes into account the EFT  corrections $\textbf{F}_{1,0},\,\textbf{F}_{0,1},\, \textbf{F}_{2,0}$ that were explicitly computed in Sec.~\ref{sec_EFT}. III is based on the minimization of the deviation of total current of the system from the expected generalized Fick's 1st law. Many higher derivative and loop corrections cancel in III, and in fact account for more corrections to leading order diffusion than II, without having to compute them explicitly. Namely, all zero-loop and one-loop higher-derivative corrections $\textbf{F}_{0,n}$ and $\textbf{F}_{1,n}$, as well as all $\ell$-loop zero-derivative corrections $\textbf{F}_{\ell,0}$ cancel in III, see Appendix~\ref{sec:fit}. However, it is a more expensive method as it requires measuring the full current which is equivalent to measuring the full structure factor. Method II, on the other hand, can be equally efficient when few points of the structure factor are sampled. Additionally, III is less accurate than II in systems with multiple conserved charges as explained in Appendix~\ref{Ap:QC}.

We now further elaborate on II, which is a general fitting scheme for the estimation of  $D(\overline{\sigma})$. II is inspired by our classical simulations (see Section~\ref{ssec_KLS}), where we found that diffusivity estimation by finite time simulations is more accurate when the corrections to the leading order diffusion are taken into account. The fitting is performed as follows: We simulate the dynamics for different equilibria magnetizations and store the local magnetization values for different sites at different times,  $s(\overline{\sigma}_i,x_j,t_k)$, where the subscripts $i,j,k$ denote different samples in the discretized dataset. The diffusivity is estimated by fitting the numerical dataset with the the EFT function Eq.~(\ref{eq:EFT_DW}) using a simple least-squares method,
\begin{equation}\label{eq_methodII}
   \text{Min}_{D,\vec{c}} \sum_{i,j,k}|s(\overline{\sigma}_i,x_j,t_k) - \mathcal{S}(\overline{\sigma}_i,x_j,t_k,D(\overline{\sigma}),\vec{c}(\overline{\sigma}))|^2,
\end{equation}
The functional minimization over $D(\overline{\sigma})$ is simplified by employing a Taylor expansion around half-filling, $D(\overline{\sigma}) = \sum^M_{i=0}b_{i}(\overline{\sigma})^{2 i}$, where $M= 3$ is found to give converged results for the parameter regimes studied in this work. Only even powers are allowed in the expansion due to the particle-hole symmetry in our system ($D(\overline{\sigma}) = D(-\overline{\sigma})$). The parameters $\vec{c} = (c_1, c_2, \ldots)$ are non-universal parameters arising from linear-fluctuations described in section ~\ref{ssec_higherderiv}. Since we will only use leading and subleading-order corrections, we just require the two parameters $\vec{c} = (c_1, c_2)$ defined by Eq.~(\ref{eq_F01}) and are present in $\textbf{F}_{0,1}$.

\subsection{Results}

\textit{Dephasing--}To establish the efficiency of the fitting methods and the accuracy of EFT predictions, we switch off the staggered field perturbation ($g=0$) and simulate the Floquet-XXZ chain, Eq.~(\ref{eq:decoherence}) in the presence of dephasing with $\gamma = 0.1$. 
Diffusive transport induced by dephasing has two distinct features. First, the single particle limit (equivalently the non-interacting limit $\Delta=0$) in the presence of dephasing remains diffusive, and therefore, diffusivity is finite for all magnetizations. This is in contrast to purely interaction-induced diffusion where the single particle limit is ballistic (free particles).  Second, for increasing strengths of dephasing, magnetization transport becomes less sensitive on the interactions in the strong noise limit where, to leading order in $1/\gamma$, $D \propto J^2 /\gamma$ \cite{bauer2017stochastic} is independent of magnetization density $\bar \sigma$ and hence $D'\equiv \partial_{\bar\sigma}D\simeq0$.  

In Figure~\ref{Fig:4}.A we show the dependence of diffusivity on magnetization, $D_{\text{III}}\sim D_{\text{II}} \approx 5.35 + 3.03 \overline{\sigma}^2 + 0.54 \overline{\sigma}^4$. For the fit II we have employed the terms $(\mathbf{F}_{0,0},\mathbf{F}_{1,0},\mathbf{F}_{0,1})$. In contrast to the non-interacting ($\Delta=0$) limit, diffusivity does have a magnetization dependence due to the presence of non-linear corrections. However, the 1-loop correction $\textbf{F}_{1,0}$ is small at timescales of order $t\sim O(100)$, since  $C_{1}\sim O(10^{-2})$. The methods II,III converge to almost the same curve (independently of bond dimension), which is indicative of the fast convergence to the asymptotic behaviour. Method I slightly underestimates the asymptotic value. The reason is that (I) does not capture the linear corrections $\textbf{F}_{0,1}$,  which dominate at these timescales for all fillings, despite being suppressed by a factor of $1/\sqrt{t}$ compared to the 1-loop corrections, Figure~\ref{Fig:4}.C. The 1-loop effects only have a visible effect at the largest simulated times. In all cases, the corrections $\Delta s$ are accurately captured from our theory. Finally, we observe that the correction profiles are almost time-independent, which is indicative that besides 1-loop corrections, higher order corrections are also suppressed. 

\textit{Unitary dynamics--} Shifting to unitary dynamics, in Figure~\ref{Fig:4}.C we show that the staggered field generates a strong dependence of diffusivity to magnetization, independently of the fitting method used, $C_{1}\propto O(1)$. Unsurprisingly, the fitted diffusivity is then considerably affected by the method used as shown by the $d=256, T= 400$ fits: $D_{\text{II}} \approx 1.77 +  13.7 \overline{\sigma}^2 + 220 \overline{\sigma}^4$, $D_{\text{III}} \approx 1.73 +  27 \overline{\sigma}^2 + 190 \overline{\sigma}^4$. For the fit II we have employed the terms $(\mathbf{F}_{0,0},\mathbf{F}_{1,0},\mathbf{F}_{0,1})$. This discrepancy is not due to truncations in the dynamics; instead, we believe it arises from fitting timescales that are not in the asymptotic diffusive regime, which is reflected by the dependence of parameters yielded by each method on the maximum timescale of the simulation. The methods are affected according to the number of corrections to diffusion they include in the approximation. In that sense, III is more converged than II, and II is more converged than I, which employs no additional corrections to asymptotic diffusion. As expected, when simulations are performed for longer times, different methods tend to show better agreement.  

Due to the discrepancy in the determination of $D$, the form of the corrections $\Delta s$ depends on the fitting method. Here we have chosen to use the diffusivity estimated by method II, since by construction it fits these corrections arising from nonlinearities and higher derivative terms, while in III the leading such effects are absent. In Appendix~\ref{Ap:QC} we show that the different methods result to corrections with similar profiles. Figure~\ref{Fig:4}.D shows that for finite magnetization $\overline{\sigma} =0.2$,  $\Delta s$ scales as $t^{-1/2}$, and has a closely matching  profile to that of the expected 1-loop correction $\mathbf{F}_{1,0}$. We again note that the 1-loop profile is completely determined by ($D,\partial_{\overline{\sigma}}D)$ and no additional fitting is involved. We observe that the correction profile shows a significant time dependence, indicative of higher corrections being still at work at these timescales. In contrast to finite equilibrium magnetization, $\overline{\sigma} =0$ requires special attention. First, the correction signal is weaker and requires a bond dimension $d = 400$ to be accurately captured. Additionally, the strength of non-linear corrections suggests that the leading corrections, $\mathbf{F}_{2,0}$  and $\mathbf{F}_{0,1}$ will be of similar magnitude at intermediate timescales. Eventually $\mathbf{F}_{2,0}$, will dominate due to its logarithmic divergence in time. For this reason we perform a different fit II around $\overline{\sigma}=0$, by including a few points $\sigma \in(0,0.01,0.02,0.03)$ and all terms $(\mathbf{F}_{0,0},\mathbf{F}_{1,0},\mathbf{F}_{0,1},\mathbf{F}_{2,0})$. Figure~\ref{Fig:4}.D indeed shows that both $\mathbf{F}_{2,0}$ and $\mathbf{F}_{0,1}$ are important at these timescales. However, the available time scales are not sufficient to see the ultimate dominance of the logarithmic part of $\mathbf{F}_{2,0}$, which would lead to a profile with an opposite sign around $y= -0.2,0.2$. 

Overall, we have shown that employing EFT corrections to study quantum transport can significantly improve the estimation of asymptotic transport parameters such as diffusivity. In addition, these corrections help understanding the different processes that drive a system towards equilibrium. For example, dephasing, which is often used to accelerate thermalization, achieves this at the cost of flattening the diffusivity as a function of filling or magnetization. We have moreover found that even if  noise $\gamma<\Delta,J$ the system's behavior is similar to the strong noise limit $\gamma\gg \Delta,J$, where the equation of motion for the conserved charge can be perturbatively derived~\cite{bauer2017stochastic,Claeys:2021skz}. In that case, the diffusivity has weak dependence on equilibrium magnetization $\overline{\sigma}$, leading to smaller loop corrections and faster thermalization: indeed the leading order non-linear terms in the diffusivity only appear at 3rd order $O(\Delta^2 J^2/\gamma^3)$. In our situation, even if perturbation theory in $1/\gamma$ is not strictly valid
, we observe the same behaviour: Weak non-linear corrections, \textbf{F}$_{1,0}$, which are almost invisible at the simulated timescales,  and a fast approach to asymptotic times which is driven by the sub-leading linear corrections \textbf{F}$_{0,1}$. 

Staggered perturbations on the other hand, induce strong non-linear effects, leading to a slower approach to equilibrium. Additionally, the classical resources required to simulate the system increase rapidly with the simulation time, and therefore, the accessible timescales are limited. We have shown that employing fitting methods which take into account the EFT corrections to diffusion leads to a significant improvement in the diffusivity estimation for $\overline{\sigma}>0$. At the same time our numerical data strongly suggest that the EFT corrections to the dynamical structure factor are present in interacting quantum systems.


\section{Discussion}\label{sec_disc}

We have employed the EFT of diffusion to derive the scaling functions of the leading power law corrections to diffusive transport for thermalizing systems with one or more conserved local charges. We confirmed these predictions by numerical simulations in a classical model, finding percent-level agreement of the entire scaling function without any fitting parameter (see Fig.~\ref{Fig:3}). While testing subleading EFT predictions in quantum simulations with this same level of precision is currently beyond reach, due to the rapid growth of required classical resources, these corrections are expected to be particularly important there due to the shorter accessible timescales. We showed that knowledge of these corrections allows for more accurate extraction of transport parameters, especially when the accessible timescales are very limited. Furthermore, our results open a number of promising directions for future research; we list these and other applications below.

\paragraph*{Precision tests of thermalization---}

Our findings can also be used to test possible deviations from standard diffusion in numerics and experiments. For example, tracking the density (or temperature) dependence of transport parameters can help estimate power law corrections to observables. Given that these corrections typically make diffusive systems appear superdiffusive at intermediate times, it would be interesting to study them quantitatively in the context of 1d chains showing apparent anomalous diffusion of superdiffusion \cite{DeNardis:2020rau,Ljubotina:2022ssg,chen2023superdiffusive} (see \cite{Glorioso:2020loc} for preliminary work in this direction, and \cite{Wienand:2023bkm} for related work), as well systems featuring subdiffusion without dipole conservation \cite{Richter:2021oxe,Gopalakrishnan:2022img}. Higher-point functions of local operators offer useful information in this regard. Indeed, we show in Appendix~\ref{sec_nonlin}, that these are controlled by the same EFT parameters (in particular, $D'$) which lead to power law corrections to linear response at intermediate times. Measuring higher-point functions of density (or heat) therefore provides a time-scale that must be exceeded to access the asymptotic dynamics. The EFT also points to other observables, such as correlation functions in momentum space $\langle n(q,t) n\rangle$, which instead are not suitable for precision tests of thermalization because they receive large fluctuation corrections \cite{Delacretaz:2020nit}.

\paragraph*{Beyond diffusion---}

Our theoretical results can be extended in various directions. We have assumed the dissipative fixed point to be diffusive, however one can similarly study corrections to subdiffusive or superdiffusive universality classes, or even in generalized hydrodynamics for integrable models (for the KPZ universality class, the leading scaling correction was studied in \cite{PhysRevLett.104.230601,PhysRevLett.104.230602,Ferrari2011}). These corrections are also important to incorporate for quantum simulations in higher dimensions $d>1$, where our ability to numerically study large systems and times is limited. Our results for the 1-loop and 2-loop corrections, Eqs.~\eqref{eq_deltaD_1loop_result}, \eqref{eq_deltaD_2loop}, hold in any dimension.

\paragraph*{Connections to simulation complexity---}

We believe that EFT corrections present new hints towards understanding the hardness of quantum simulations in the linear response regime. We found that classical resources increase with time faster when non-linear corrections are stronger, in our case when $\overline{\sigma}$ increases. This obstructed exploring magnetizations beyond $\overline{\sigma}\approx 0.25$ in the staggered field simulations. We lack a detailed theory behind this observation, but believe that this is related to the strong non-linear corrections, since they enhance multi-point correlation functions such as the ones explored in Appendix~\ref{sec_nonlin}. This implies that the accurate simulation of a  system with strong non-linear contributions requires keeping more information on multi-body correlations in the density matrix, which in turn increases the resources (bond dimension) required by the tensor network simulations.

\paragraph*{Benchmark for new methods---}
Our results on universal corrections to hydrodynamics are also useful to benchmark theoretical and computational~\cite{RafaelPRB18,Parker19PRX,Claeys:2021skz,klein2022time,rakovszky2022dissipation} approaches to thermalization in many-body systems, as these will have to reproduce not only the leading diffusive behavior but the corrections as well. For example, Ref.~\cite{Parker19PRX} approximates correlation functions based on extrapolations of Lanczos coefficients, which by design produces a meromorphic $G^R(\omega,q)$ that cannot capture the universal non-analytic corrections \eqref{eq_deltaD_1loop_d1}. Incorporating EFT results into such constructions is a promising path to `bootstrapping' transport in correlated quantum systems.

\paragraph*{Distinguishing theories of fluctuating hydrodynamics---}
We have also shown that high precision classical stochastic simulations offer valuable precision tests for theories of fluctuating hydrodynamics, in the present case confirming the leading and subleading corrections predicted by the Schwinger-Keldysh EFT approach \cite{Crossley:2015evo,Chen-Lin:2018kfl}. Other approaches for fluctuating hydrodynamics exist, which treat the noise fields somewhat differently; it would be interesting to further push these tests, to possibly rule out certain theories and identify the correct systematic framework. One possible concrete target for the numerics in this regard are effects arising from non-Gaussianities in noise fields that do not enter in constitutive relations \cite{Jain:2020hcu}.

\section*{Acknowledgements}

We thank Paolo Glorioso and Zlatko Papic for insightful discussions. This work was partially supported  by the European Research Council via the grant agreements TANQ 864597 (A.M., D.A.) and by the Swiss National Science Foundation (D.A.). 

\section*{Appendix}
\appendix

\section{EFT details}\label{ap:A}
\subsection{General structure of the corrections}\label{ap:A1}

The corrections to diffusion shown in Eq.~\eqref{eq_nn_noph} and \eqref{eq_nn_ph} correspond to the first few terms arising from an expansion in fluctuations and derivatives in the EFT. The general form that this expansion takes, in $d$ spatial dimensions, is:
\begin{widetext}
\begin{equation}\label{eq_correlator_generalform}
\begin{split}
\langle n(x,t)n\rangle
	= \frac{\chi}{(4\pi D t)^{d/2}} 
	\bigg[\ \ &\ F_{0,0}(y) + \frac{1}{t} F_{0,1}(y) + \frac{1}{t^2} F_{0,2}(y) +\cdots \\
+ \frac{1}{t^{d/2}} &\left( F_{1,0}(y) + \frac{1}{t} F_{1,1}(y) + \frac{1}{t^2} F_{1,2}(y) +\cdots \right)\\
+ \frac{1}{t^{d}} &\left( F_{2,0}(y) + \frac{1}{t} F_{2,1}(y) + \frac{1}{t^2} F_{2,2}(y) +\cdots \right) +\cdots \ \bigg] \, ,
\end{split}
\end{equation}
\end{widetext}
where $F_{\ell,n}$ are scaling functions of the scaling variable $y\equiv x/\sqrt{Dt}$. The overall form of the expansion \eqref{eq_correlator_generalform} is simple to justify on general grounds: higher derivative corrections to diffusion come with two derivatives (assuming reflection or rotation symmetry), and therefore give corrections suppressed by $\nabla^2\sim \frac{1}{x^2}\sim \frac{1}{t}$ at late times. Loop corrections instead come from nonlinearities in the dynamics of the densities: a single cubic nonlinearity is suppressed by $\delta n \sim q^{d/2} \sim 1/t^{d/4}$ compared to the linear (Gaussian) dynamics. The first loop correction requires two insertions of a cubic nonlinearity, and is hence $1/t^{d/2}$ suppressed. Generalizing, an $\ell$-loop contributions at $n$th order in the derivative expansion will give a correction to correlation functions suppressed by $1/t^{n+\ell d/2}$ (up to logarithms); this correction comes with a dimensionless scaling function $F_{\ell,n}$ and is shown in the $\ell$th line and $n$th column in Eq.~\eqref{eq_correlator_generalform}. This general structure of corrections to hydrodynamics applies not only to density two-point functions, but also to higher point functions \cite{Delacretaz:2023ypv}, as well as to correlators of arbitrary microscopic operators that have the same quantum numbers as (composites of) densities \cite{Delacretaz:2020nit,Glorioso:2020loc}.

While the simple scaling argument above predicts the general expansion of correlation functions at late times in diffusive systems, obtaining the dimensionless scaling functions $F_{\ell,n}$ in Eq.~\eqref{eq_correlator_generalform} requires detailed use of the EFT. The leading diffusive scaling function is well known $F_{0,0}(y) = e^{-y^2/4}$, and captures the density two-point function universally in any diffusive system. The subsequent $F_{\ell,n}$ capture scaling corrections to diffusion; they are also universal, up to one or a few theory-dependent factors.

In this paper, we focus on the first few corrections and explicitly evaluate $F_{1,0}$ (Eq.~\eqref{eq_F10}), $F_{2,0}$ (Eq.~\eqref{eq_F20}), and $F_{0,1}$ (Eq.~\eqref{eq_F01}). Note that in \eqref{eq_correlator_generalform}, we have suppressed certain factors of $\log t$ that can arise from loop corrections coming with integer powers of $1/t$, see \eqref{eq_F20} for an example.

\subsection{Details of the EFT}\label{app_EFTdetails}

In this section, we further motivate the EFT representation of the generating functional \eqref{eq_Z_EFT}, and go over several key steps in the construction of the EFT. Most of the discussion in this section can be found elsewhere, e.g.~Refs.~\cite{Crossley:2015evo,kamenev2023field}, but we include it for completeness.

One of the guiding principles in constructing to the Schwinger-Keldysh EFT for hydrodynamics is to introduce a minimal set of fluctuating degrees of freedom that will ensure gauge invariance of the generating functional
\begin{equation}\label{eq_Z_gauge}
Z[A^1_\mu + \partial_\mu \lambda^1, A^2_\mu + \partial_\mu \lambda^2]
	= Z[A^1_\mu, A^2_\mu] \, .
\end{equation}
This is achieved by introducing phases $\phi^{1,2}$ that always enter in the combination $A_\mu + \partial_\mu \phi$ (sometimes called the `St\"uckelberg trick'):
\begin{equation}\label{eq_Z_L_app}
Z[A^1_\mu, A^2_\mu]
	= \int D\phi^1 D \phi^2 \, e^{i \int dt d^dx \mathcal L[A_\mu^1 + \partial_\mu \phi^1, A_\mu^2 + \partial_\mu \phi^2]} \, .
\end{equation}
This now satisfies \eqref{eq_Z_gauge} for any functional $\mathcal L$, because a gauge transformation can be absorbed through a redefinition of the dynamical fields $\phi^{1,2}$ that are being integrated over. It is clear that the degrees of freedom we have introduced are related to the continuous symmetry of the system. If one had considered instead a system with $N$ separate continuity relations \eqref{eq_continuityrel}, $2N$ fields would have been introduced. One already notices a ressemblence with earlier approaches to fluctuating hydrodynamics, where each continuity relation leads to two degrees of freedom: a density and an associated noise field. The central assumption in the construction of the EFT is that $\mathcal L$ is a local functional of the fields $\phi^{1,2}$. This implements the expectation of thermalization: the only long-lived quantities are associated with symmetries, so that integrating out other degrees of freedom produces a local EFT with a derivative expansion controlled by the scale at which the system thermalizes (this scale acts as the UV cutoff of the hydrodynamic EFT).

It is useful to define the symmetric and antisymmetric combinations of fields \cite{kamenev2023field}
\begin{equation}
\phi_r \equiv \frac{\phi_1+\phi_2}{2}\, , \qquad
\phi_a \equiv {\phi_1-\phi_2}\, .
\end{equation}
One advantage of fields in this basis is that they satisfy the `latest time' property: correlators where the latest time is carried by a $\phi_a$ field vanish due to cyclicity of the trace in Eq.~\eqref{eq_Z}
\begin{equation}\label{eq_latesttime}
\langle O_1(t_1)\cdots O_n(t_n) \phi_a(t_{n+1})\rangle=0\, , \quad t_{n+1}>t_i\, ,
\end{equation}
which will lead to simplifications in diagrams below.

The construction bears a resemblance with EFTs for spontaneously broken phases, where the long-lived degrees of freedom are Goldstone bosons. In fact, constructing the most general local Lagrangian $\mathcal L$ in \eqref{eq_Z_L_app} leads to an effective description of a thermalizing system in the symmetry broken phase (a dissipative superfluid). To describe the normal phase, Ref.~\cite{Crossley:2015evo} proposes to forbid the propagating sound mode by imposing an additional symmetry:
\begin{equation}
\phi_r(x,t) \to \phi_r(x,t) + \lambda(x)\, .
\end{equation}
See Refs.~\cite{deBoer:2018qqm,Glorioso:2018mmw} for discussions on this symmetry in a holographic context. Recently, Ref.~\cite{Akyuz:2023lsm} proposed a slightly different approach that bypasses the need to impose this somewhat artificial symmetry, by viewing the density $n_r$ rather than $\phi_r$ as the fundamental degree of freedom of the EFT. We expect both of these approaches to be equivalent.

Otherwise, one simply constructs the most general local functional of the gauge invariant combinations of fields
\begin{equation}
B_\mu^1 \equiv A_\mu^1 + \partial_\mu \phi^1\, , \qquad
B_\mu^2 \equiv A_\mu^2 + \partial_\mu \phi^2\, , 
\end{equation}
in an expansion in fields and derivatives. There are few additional constraints to impose, such as unitarity $Z[A_1,A_2]^*=Z[A_2,A_1]$ (which simply follows from the definition \eqref{eq_Z}), and KMS symmetry; we refer the reader to Ref.~\cite{Crossley:2015evo} for details. To leading order in derivatives, the action can be expressed as \cite{Delacretaz:2023ypv}
\begin{equation}\label{eq_L_n}
\mathcal L = 
	\sigma(n) B_{ai} \left( i B_{ai} - \beta E_{ri}\right)
	+ B_{a0} n - D(n)B_{ai}	\partial_i n + \cdots\, ,
\end{equation}
where $\beta$ is the inverse temperature, $E_{ri} = \partial_0 A_{ri} - \partial_i A_{r0} $ is the electric field, and $B_{a\mu} = A_{a\mu} + \partial_\mu \phi_a$. We have changed variables from $\phi_r$ to the density $n$. The ellipses denote higher derivative terms (the most important of which are discussed separately in Sec.~\ref{ssec_higherderiv}), as well as nonlinear terms that contain higher powers of $\phi_a$, which are more irrelevant than the nonlinearities considered here.

Setting the background fields to zero $A^{1,2}\to 0$ leads to the action \eqref{eq_L_EFT} used in the main text. The background fields are however useful to generate various correlation functions. For example, the retarded Green's function of charge density is
\begin{equation}
\begin{split}
G^R(t,x)
	&\equiv i \theta(t) {\rm Tr} \left(\rho [n(t,x),n]\right)\\
	&= i \langle n_r(t,x) n_a\rangle\\
	&=i\frac{ \delta^2 \log Z}{\delta (iA_{a0}(t,x)) \delta (i A_{r0}(0))}\\
	&= -i \langle n(t,x) \partial_i \left(\sigma(n)\partial_i \phi_a\right)\rangle\, .
\end{split}
\end{equation}
In particular, when $\sigma(n) = \sigma = {\rm const}$, the retarded Green's function is simply related to the $\langle n \phi_a\rangle$ propagator
\begin{equation}\label{eq_GR_nphia_simple}
G^R(\omega,q)
	= i\sigma q^2 \langle n\phi_a\rangle(\omega,q)\, .
\end{equation}
The two-point function can be obtained from $G^R$ as usual from a fluctuation-dissipation relation
\begin{equation}\label{eq_FDR}
\begin{split}
\langle nn\rangle(\omega,q)
	&= \frac{2}{1-e^{-\beta\omega}} \Im G^R(\omega,q)\, . 
\end{split}
\end{equation}
%

\subsection{1-loop calculation}\label{ap:A3}

The universal leading 1-loop correction to diffusion was computed in Ref.~\cite{Chen-Lin:2018kfl} (see also Refs.~\cite{Jain:2020zhu,Jain:2020hcu,Winer:2020gdp,Sogabe:2021svv,Abbasi:2022aao,Chao:2023kvz} for further studies of loop effects in the EFT of diffusion). We review the derivation here, and discuss an interesting cancellation between certain diagrams that simplifies the calculation. 

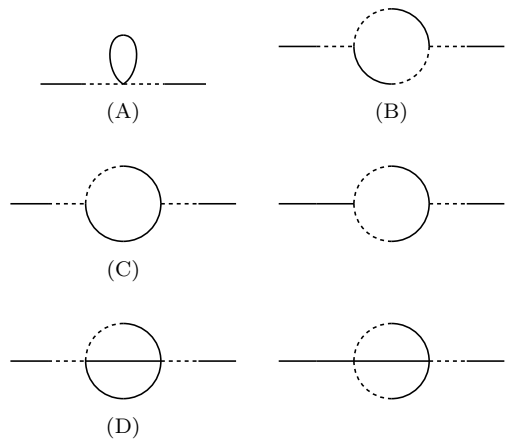
\begin{figure}[h] 
\centerline{
	\begin{subfigure}{.4\linewidth}
	\centering
	\scalebox{0.5}{
		\begin{tikzpicture}
		\begin{feynman} \vertex at (0,0) (i1);
		\vertex at (1.1,0) (i2);
		\vertex at (2.2,0) (i3);
		\vertex at (2.2,1.3) (I3);
		\vertex at (3.3,0) (i4);
		\vertex at (4.4,0) (i5);
		\diagram*{
			(i1) -- [plain, very thick] (i2),
			(i2) -- [scalar, very thick] (i3),
			(i3) -- [plain, very thick, out=150, in=180] (I3),
			(i3) -- [plain, very thick, out=30, in=0] (I3),
			(i3) -- [scalar, very thick] (i4),
			(i4) -- [plain, very thick] (i5)
		};
		\end{feynman}
		\end{tikzpicture}
	}\caption{\label{sfig_tad}}
	\end{subfigure}
	\begin{subfigure}{.4\linewidth}
	\centering
	\scalebox{0.5}{
		\begin{tikzpicture}
		\begin{feynman} \vertex at (0,0) (i1);
		\vertex at (1,0) (i2);
		\vertex at (2,0) (i3);
		\vertex at (3,1) (i4);
		\vertex at (3,-1) (I4);
		\vertex at (4,0) (i5);
		\vertex at (5,0) (i6);
		\vertex at (6,0) (i7);
		\diagram*{
			(i1) -- [plain, very thick] (i2),
			(i2) -- [scalar, very thick] (i3),
			(i3) -- [scalar,quarter left, very thick] (i4),
			(i4) -- [plain,quarter left, very thick] (i5),
			(i3) -- [plain,quarter right, very thick] (I4),
			(I4) -- [scalar,quarter right, very thick] (i5),
			(i5) -- [scalar, very thick] (i6),
			(i6) -- [plain, very thick] (i7)
		};
		\end{feynman}
		\end{tikzpicture}
	}\caption{\label{sfig_spsp}}
	\end{subfigure}
}
	\par\bigskip
\centerline{
	\begin{subfigure}{.4\linewidth}
	\centering
	\scalebox{0.5}{
		\begin{tikzpicture}
		\begin{feynman} \vertex at (0,0) (i1);
		\vertex at (1,0) (i2);
		\vertex at (2,0) (i3);
		\vertex at (3,1) (i4);
		\vertex at (3,-1) (I4);
		\vertex at (4,0) (i5);
		\vertex at (5,0) (i6);
		\vertex at (6,0) (i7);
		\diagram*{
			(i1) -- [plain, very thick] (i2),
			(i2) -- [scalar, very thick] (i3),
			(i3) -- [scalar,quarter left, very thick] (i4),
			(i4) -- [plain,quarter left, very thick] (i5),
			(i3) -- [plain,quarter right, very thick] (I4),
			(I4) -- [plain,quarter right, very thick] (i5),
			(i5) -- [scalar, very thick] (i6),
			(i6) -- [plain, very thick] (i7)
		};
		\end{feynman}
		\end{tikzpicture}
	}\caption{\label{sfig_spDp}}
	\end{subfigure}
	\begin{subfigure}{.4\linewidth}
	\centering
	\scalebox{0.5}{
		\begin{tikzpicture}
		\begin{feynman} \vertex at (0,0) (i1);
		\vertex at (1,0) (i2);
		\vertex at (2,0) (i3);
		\vertex at (3,1) (i4);
		\vertex at (3,-1) (I4);
		\vertex at (4,0) (i5);
		\vertex at (5,0) (i6);
		\vertex at (6,0) (i7);
		\diagram*{
			(i1) -- [plain, very thick] (i2),
			(i2) -- [plain, very thick] (i3),
			(i3) -- [scalar,quarter left, very thick] (i4),
			(i4) -- [plain,quarter left, very thick] (i5),
			(i3) -- [scalar,quarter right, very thick] (I4),
			(I4) -- [plain,quarter right, very thick] (i5),
			(i5) -- [scalar, very thick] (i6),
			(i6) -- [plain, very thick] (i7)
		};
		\end{feynman}
		\end{tikzpicture}
	}\caption*{}
	\end{subfigure}
}
	\par\bigskip
\centerline{
	\begin{subfigure}{.4\linewidth}
	\centering
	\scalebox{0.5}{
		\begin{tikzpicture}
		\begin{feynman} \vertex at (0,0) (i1);
		\vertex at (1,0) (i2);
		\vertex at (2,0) (i3);
		\vertex at (3,1) (i4);
		\vertex at (3,-1) (I4);
		\vertex at (4,0) (i5);
		\vertex at (5,0) (i6);
		\vertex at (6,0) (i7);
		\diagram*{
			(i1) -- [plain, very thick] (i2),
			(i2) -- [scalar, very thick] (i3),
			(i3) -- [scalar,quarter left, very thick] (i4),
			(i4) -- [plain,quarter left, very thick] (i5),
			(i3) -- [plain,quarter right, very thick] (I4),
			(I4) -- [plain,quarter right, very thick] (i5),
			(i3) -- [plain, very thick] (i5),
			(i5) -- [scalar, very thick] (i6),
			(i6) -- [plain, very thick] (i7)
		};
		\end{feynman}
		\end{tikzpicture}
	}\caption{\label{sfig_sppDpp}}
	\end{subfigure}
	\begin{subfigure}{.4\linewidth}
	\centering
	\scalebox{0.5}{
		\begin{tikzpicture}
		\begin{feynman} \vertex at (0,0) (i1);
		\vertex at (1,0) (i2);
		\vertex at (2,0) (i3);
		\vertex at (3,1) (i4);
		\vertex at (3,-1) (I4);
		\vertex at (4,0) (i5);
		\vertex at (5,0) (i6);
		\vertex at (6,0) (i7);
		\diagram*{
			(i1) -- [plain, very thick] (i2),
			(i2) -- [plain, very thick] (i3),
			(i3) -- [scalar,quarter left, very thick] (i4),
			(i4) -- [plain,quarter left, very thick] (i5),
			(i3) -- [scalar,quarter right, very thick] (I4),
			(I4) -- [plain,quarter right, very thick] (i5),
			(i3) -- [plain, very thick] (i5),
			(i5) -- [scalar, very thick] (i6),
			(i6) -- [plain, very thick] (i7)
		};
		\end{feynman}
		\end{tikzpicture}
	}\caption*{}
	\end{subfigure}
}
\caption{\label{fig_feyndiag_appendix} Diagrams {\em not} contributing to transport corrections to diffusion: {\bf (a)} Diagrams where the external momentum does not flow through a loop cannot produce new IR singularities; they only renormalize tree-level transport parameters. {\bf (b)} The 1-loop contribution proportional to $\sigma'^2$ vanishes due to the latest time condition \eqref{eq_latesttime}. {\bf (c)} The two 1-loop contributions proportional to $\sigma' D'$ cancel. {\bf (d)} A similar cancellation happens at 2-loop for the $\sigma'' D''$ contribution. }
\end{figure}

We focus on non-analytic and UV finite corrections to $G^R(\omega,q)$. This correlator also receives UV divergent corrections, which can be absorbed with local counterterms in the EFT, and renormalize existing transport parameters. These can be interesting in their own right \cite{Kovtun:2011np}, but do not lead to power-law corrections to diffusive behavior, which are the focus of this paper. We can therefore omit diagrams such as the one shown in Fig.~\ref{sfig_tad}, which cannot produce novel singular IR structure. One therefore only needs to  the cubic vertices of the EFT:
\begin{equation}\label{eq_L3_app}
\mathcal L^{(3)}
	= i\sigma' n (\nabla \phi_a)^2 + \frac12 D' \nabla^2 \phi_a n^2\, .
\end{equation}
Diagramatically, one expects these will generate 1-loop corrections proportional to $\sigma'^2$, $\sigma'D'$, and $D'^2$. The first is shown in Fig.~\ref{sfig_spsp} and can easily be seen to vanish: indeed, the only such diagram involves a loop where all poles in frequency lie on the same half complex plane. Alternatively, this can be seen to vanish in time domain using the latest time condition \eqref{eq_latesttime}. Indeed, the diagram involves the computation of a correlator $\langle (n\phi_a)(t_1) (n\phi_a)(t_2)\rangle$ -- since a $\phi_a$ field appears at the latest time (whether it is $t_1$ or $t_2$), the correlator must vanish. These types of considerations were already well known to produce diagrammatic simplifications in Schwinger-Keldysh EFTs, see, e.g., Ref.~\cite{Gao:2018bxz}. There is in fact a more general argument, showing that the EFT \eqref{eq_L_EFT} with $D(n) = \rm const$ but $\sigma(n)$ arbitrary has no loop corrections to the two-point function; acting with the diffusive kernel on the two point function, one has
\begin{equation}\label{eq_eom_trick}
(\partial_t - D \nabla^2) \langle n(t,x)n\rangle
	= -2i\nabla \langle \left(\sigma(n)\nabla\phi_a\right)_{t,x} n\rangle\, ,
\end{equation}
where we used the equation of motion $\delta \mathcal L /\delta \phi_a = 0$ (this equation holds up to contact terms $\propto \delta(t)\delta (x)$. The right-hand side vanishes because a $\phi_a$ field is at the latest time (assuming without loss of generality that $t>0$), showing that the two-point function is unaffected by nonlinearities in this theory, and is equal to
\begin{equation}
\langle n(t,x)n\rangle = \frac{\chi}{(4\pi Dt)^{d/2}} e^{-x^2/(4Dt)} \, .
\end{equation}
This is known to occur in certain lattice gas models \cite{spohn2012large}. Here we have temporarily ignored higher-derivative corrections, which will enter as usual through $F_{0,n}$ as in \eqref{eq_correlator_generalform}.

We have established that the $\sigma'^2$ contribution to $\langle nn\rangle(\omega,q)$ vanishes. One can in fact show that the $\sigma' D'$ contribution vanishes as well, although the argument is slightly more subtle. Diagramatically, the two diagrams that give corrections to $\langle nn\rangle(\omega,q)$ are shown in Fig.~\ref{sfig_spDp}. They can be shown to cancel by explicit calculation---however, the cancellation only happens after performing the integral over frequency, and dropping a UV divergence in the integral over momenta. Note that, after amputating one external leg on the $D'$ vertex, the remaining object to be computed is a two-point function between $n$ and the normal-ordered composite operator $n^2$:
\begin{equation}
\langle n(t,x) n^2(0,0)\rangle\, .
\end{equation}
Crucially, this object is to be computed in the theory with $\sigma'$ as its only cubic interaction (since the $D'$ vertex has already been used). By time-reversal symmetry, one can take $t>0$. Acting with the diffusive kernel and using the equation of motion as in \eqref{eq_eom_trick}, one finds again that the result vanishes, which implies that this correlator must be proportional to the diffusive two-point function
\begin{equation}
\langle n(t,x) n^2(0,0)\rangle
	\propto \frac{1}{t^{d/2}}e^{-x^2/(4Dt)}\, .
\end{equation}
The diagram therefore at most renormalizes $\chi$ or $D$, without producing new non-analytic structures.

We are left with the $D'^2$ contribution to the two-point function, shown in Fig.~\ref{fig_12loop}. It is simplest to study the 1-loop correction to the retarded Green's function $G^R$, which is simply related to $\langle n\phi_a\rangle(\omega,q)$ by \eqref{eq_GR_nphia_simple} (given that we have shown that one can set $\sigma(n) = \sigma = $ const in the action). This is done in Sec.~\ref{sec_EFT}, where it is shown that the loop can be expressed as a correction $D\to D+ \delta D(\omega,q)$ with \eqref{eq_deltaD_1loop}
\begin{align}\label{eq_app_deltaD_1loop}
\delta D(\omega,q)
	&= -i D'^2 \int_{p'} q'^2 \langle n\phi_a\rangle(p') \langle nn\rangle(p-p')\\
	&= -i \chi (D')^2 \int \frac{d^d q'}{(2\pi)^d} \frac{q'^2}{\omega+ i D \left[q'^2 + (q-q')^2\right]}\notag
\end{align}
where in the second line we inserted the propagators \eqref{eq_propagators} and evaluated the integral over frequencies $\int \frac{d\omega}{2\pi}$. Here $\chi\equiv \sigma/D$. Changing integration variables to $q'\to k \equiv q'-\frac12 q$ and defining
\begin{equation}
z \equiv q^2 - \frac{2i\omega}{D}\, , 
\end{equation}
this becomes
\begin{equation}
\begin{split}
\delta D(\omega,q)
	&= - \frac{\chi D'^2}{2 D} \int \frac{d^d k}{(2\pi)^d}\\
	&-i\omega \frac{ \chi D'^2}{D^2} \int \frac{d^d k}{(2\pi)^d} \frac{1}{z + 4 k^2}\, .
\end{split}
\end{equation}
The term in the first line is a UV-divergent contribution to the diffusivity, and can be absorbed with a counterterm $D\to D + \delta D$ in the EFT. The term in the second line instead has interesting non-analytic IR behavior. It is entirely UV finite in $d=1$. In $d\geq 2$ it produces additional UV divergences that are analytic in $z$, and can be absorbed by higher-derivative counterterms in the EFT. The UV finite part is given by
\begin{equation}
\begin{split}
\alpha_d(z)
	&\equiv
\int \frac{d^d k}{(2\pi)^d}
	\frac{1}{z+ 4 k^2} - {\rm UV}\\
	&= \frac{(-z)^{\frac{d}2-1}}{(16\pi)^{d/2}\Gamma (\frac{d}{2})}  \cdot
\begin{cases}
i\pi	& \hbox{if $d$ odd}\, ,\\
\log\frac1{z}  		& \hbox{if $d$ even} \, .
\end{cases}
\end{split}
\end{equation}
One therefore finds Eq.~\eqref{eq_deltaD_1loop_result}.

\subsection{2-loop calculation}\label{ap:2loop}
In systems with charge conjugation symmetry, the EFT must by invariant under
\begin{equation}
n\to -n\, , \qquad
\phi_a\to -\phi_a\, .
\end{equation}
This forbids cubic terms in the EFT: in particular, $D',\,\sigma'=0$. The leading nonlinearities are instead quartic, and can be found again by expanding \eqref{eq_L_EFT}:
\begin{equation}
\mathcal L^{(4)}
	= \frac{i}2\sigma'' n^2 (\nabla \phi_a)^2 + \frac16 D'' \nabla^2 \phi_a n^3 \, .
\end{equation}
The leading fluctuation correction to transport then comes from 2-loop diagrams involving the two quartic vertices, such as those in Figs.~\ref{fig_12loop} and \ref{sfig_sppDpp}. These have not been computed before -- we evalute them below. The contribution proportional to $\sigma''^2$ vanishes due to the latest time condition \eqref{eq_latesttime}. The contributions proportional to $\sigma'' D''$ can also be shown to vanish, following the same argument as in the previous section: they involve computing $\langle n(t,x)n^3(0,0)\rangle$ in the theory with only the quartic interaction  $\sigma''$. The only remaining contribution is the one proportional to $D''^2$. It is studied in Sec.~\ref{sec_EFT} and leads to a correction (see Eq.~\eqref{eq_deltaD_2loop})
\begin{align}
&\delta D(\omega,q)\notag\\ \notag
&= - \frac{i}{2}D''^2\!\!\int_{p',p''}\!\!\!q''^2 \langle n \phi_a\rangle(p'') \langle nn\rangle(p'-p'') \langle nn\rangle(p-p')\\
&=
	\frac{\chi D''^2}{2D^2} \int_{p'} (-i\omega' )\alpha_d \left(q'^2 - \frac{2i\omega'}{D}\right) \langle nn\rangle(p-p')\\ \notag
&= \frac{(\chi D'')^2}{2D^2} \times\\
&\int_{q'}\!\! \left(-i\omega + D(q-q')^2\right) \alpha_d \left(q'^2 +2(q-q')^2 - \frac{2i\omega}{D}\right)\, ,\notag
\end{align}
where in the third line we used the result of the 1-loop calculation \eqref{eq_app_deltaD_1loop}, and in the last one we performed the $\omega'$ integral by residue (note that $\alpha_d(q'^2-\frac{2i\omega'}{D})$ is analytic in the upper-half $\omega'$ plane). Changing integration variables to $q'\to k\equiv \frac3{\sqrt{2}}(q'-\frac23q)$ leads to 
\begin{equation}
\begin{split}
\delta D&(\omega,q)
	= \frac{(\chi D'')^2}{2D^2} \left(\tfrac{\sqrt{2}}{3}\right)^d \times \\
	&\int_{k}\!\! \left(-i\omega + \frac19D(q^2 + 2 k^2)\right) \alpha_d \left(\frac23(k^2 + z)\right)\, , 
\end{split}
\end{equation}
where we defined 
\begin{equation}
z\equiv q^2 - \frac{3i\omega}{D}\, .
\end{equation}
Let us first focus on $d$ odd, where one can write 
\begin{equation}\label{eq_alphad_ad}
\alpha_d(z)
	= a_d z^{\frac{d}2-1}\, ,
\end{equation}
with $a_d = \frac{(-1)^{\frac{d-1}2}\pi}{(16\pi)^{d/2}\Gamma(d/2)}$. The corrections then takes the form
\begin{equation}
\begin{split}
\delta D(\omega,q)
	&= \frac{3(\chi D'')^2}{4 D^2}\times\\
	& \left[\left(-i\omega + \frac{D}{9}q^2\right)J_{d,0}(z) + \frac{2D}{9} J_{d,2}(z)\right]\, , 
\end{split}
\end{equation}
with 
\begin{equation}
J_{d,n}(z)
	= \left(\tfrac{\sqrt{2}}{3}\right)^d a_d \int \frac{d^dk}{(2\pi)^d} k^n (k^2 + z)^{\frac{d}2-1}\, .
\end{equation}
This final integral has several power-law UV divergences: these are analytic in $\omega,q$ and can therefore be absorbed with counterterms in higher-derivative corrections to hydrodynamics. We thus focus on the UV finite (or UV log-divergent), non-analytic part. By dimensional analysis, it has the form
\begin{equation}
J_{d,n}(z)
	= z^{d-1+\frac{n}2} \left[b_{d,n}\log \frac{\Lambda^2}{z} + c_{d,n}\right]  + \hbox{UV}\, .
\end{equation}
Note that the $c_{d,n}$ contribution is always analytic, so that it can be ignored. The coefficients  of interest are found to be
\begin{equation}
b_{d,0}
	= \frac{1}{(12\sqrt{3}\pi)^d} \frac{1}{\Gamma(d)}\, , \qquad
b_{d,2}
	= -\frac12 b_{d,0}\, .
\end{equation}
This second equality implies that the $q^2 z^{d-1}\log\frac1z$ correction vanishes, so that $\delta D(\omega,q)$ is proportional to $(-i\omega) z^{d-1}\log\frac1z$ -- this guarantees that static correlators $\omega\to 0$ are analytic as expected \cite{Jain:2020hcu}. We are then left with
\begin{equation}\label{eq_deltaDwq_twoloop}
\delta D(\omega,q)
	= \frac{1}{2} b_{d,0} \frac{(\chi T D'')^2}{D^2} (-i\omega) z^{d-1}\log \frac{1}{z}\, .
\end{equation}
One can repeat this calculation in $d$ even, where instead of \eqref{eq_alphad_ad} one has $\alpha_d(z) = \tilde a_d z^{\frac{d}2-1}\log\frac1z$. One finds the same results as above, apart for a sign $b_{d,n}\to -b_{d,n}$. The result in general dimension therefore takes the form \eqref{eq_deltaD_2loop_result}.

\subsection{Fourier transformation of corrections to diffusion}\label{ap:FT}

In this section, we detail the computation of the inverse Fourier transform that produces the two-point function, focusing on $d=1$ for simplicity
\begin{equation}
\langle n(t,x)n\rangle
	\equiv \int \frac{d\omega dq}{(2\pi)^2} e^{-i\omega t + i q x}
	\langle nn\rangle(\omega,q)\, ,
\end{equation}
where $\langle nn\rangle(\omega,q)$ can be obtained from the retarded Green's function using \eqref{eq_FDR}
\begin{equation}
\begin{split}
\langle nn\rangle(\omega,q)
	&\simeq \frac{2}{\omega} \Im  \frac{\sigma q^2}{-i\omega + (D + \delta D(\omega,q))q^2}\\
	&\simeq \frac{2\sigma q^2}{\omega^2 + D^2q^4}  - \frac{2}{\omega}\Im \frac{\sigma q^4 \delta D(\omega,q)}{(-i\omega+ D q^2)^2}\, , 
\end{split}
\end{equation}
where we have expanded to linear order in $\delta D$ (given by \eqref{eq_deltaD_1loop_d1} or \eqref{eq_deltaD_2loop_d1}) because we are only interested in the leading correction.

The Fourier transformation of the first term is straightforward, and given by
\begin{equation}
\langle n(t,x)n\rangle_{0}
	= \frac{\chi}{\sqrt{4\pi D t}} e^{-y^2/4}\, , 
	\qquad y\equiv \frac{x}{\sqrt{Dt}} \,.
\end{equation}
Let us now turn to the Fourier transform of the correction. Introducing the dimensionless variables $w=\omega/(Dq^2)$ and $\tau = t D q^2$, one has
\begin{equation}\label{eq_FT_tmp}
\begin{split}
\delta\langle nn\rangle(t,q)
	&= \int \frac{d\omega}{2\pi} e^{-i\omega t} \langle nn\rangle(\omega,q)\\
	&= {\chi q^2} \int \frac{dw}{2\pi}e^{-iw\tau} \frac{\delta D/(-i\omega)}{(1-iw)^2}\, .
\end{split}
\end{equation}
We have taken $t>0$, implying that one can disregard non-analyticities in the upper-half $w$ plane. Inserting the 1-loop expression \eqref{eq_deltaD_1loop_d1} leads to 
\begin{equation*}
\delta\langle nn\rangle(t,q) =
\frac{i\chi^2 D'^2|q|e^{-\tau/2}}{4\sqrt{2} D^2} \! \int_{-\infty}^0\! \frac{dz}{2\pi} {\rm Disc} \frac{e^{z\tau } }{\sqrt{z}(z+\frac12)^2}\, ,
\end{equation*}
where we defined $z=\frac12-iw$ and deformed the contour to pick up the discontinuity across the two-diffuson branch cut, ${\rm Disc}f(z) = f(z+i0^+) - f(z-i0^+)$. Evaluating the integral gives
\begin{equation}
\delta\langle nn\rangle(t,q)
	= \frac{\chi^2 D'^2|q|e^{-\tau/2}}{4 D^2 } \left[e^{-\tau/2}(1+\tau) {\rm Erfi}\!\left(\tfrac{\sqrt{\tau}}{\sqrt2}\right)-\frac{\sqrt{2\tau}}{\sqrt\pi}\right]
\end{equation}
with ${\rm Erfi}(z) \equiv {\rm Erf}(iz)/i$. To perform the final Fourier transform $\int \frac{dq}{2\pi} e^{iqx}$, one can express the integrand as a product of two Fourier transforms, and evaluate their convolution. This gives Eq.~\eqref{eq_nn_noph}.

\begin{figure}
	\centering
	\includegraphics[width=0.4\textwidth]{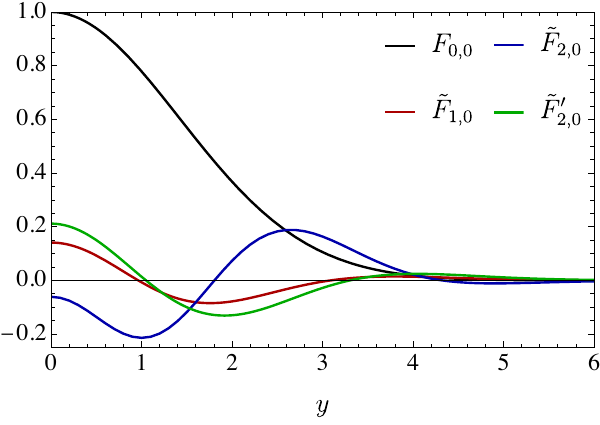}
	\caption{\label{Fig:scaling_functions} Universal scaling functions describing non-linear corrections to the diffusive structure factor as a function of the hydrodynamic variable $y = x/\sqrt{D t}$. The leading order Gaussian spreading is $F_{0,0} = e^{-y^2/4}$ and the leading order non-linear fluctuations are given by Eq.~(\ref{eq_F10}) for general systems and by Eq.~(\ref{eq_F20}) in the presence of particle-hole symmetry.}
\end{figure}

The 2-loop contribution can be obtained similarly: one inserts \eqref{eq_deltaD_2loop_d1} in \eqref{eq_FT_tmp}, and evaluates the discontinuity across the 3-diffuson branch cut to obtain
\begin{equation*}
\delta \langle nn\rangle (t,q)
	=  \frac{\chi^3 D''^2q^2}{24\sqrt{3}\pi D^2}  \left[\tau e^{-\tau} \left(\log \tfrac{1}{q^2}+{\rm Ei}(\tfrac{2\tau}{3})\right) - \frac32 e^{-\frac{\tau}{3}}\right]
\end{equation*}
where ${\rm Ei}(z)\equiv -\int_{-z}^\infty \frac{du}{u} e^{-u}$ ({\sf ExpIntegralEi[z]} in Mathematica). We were not able to express the final Fourier transform $\int \frac{dq}{2\pi} e^{iqx}$ in terms of known special functions -- the resulting integral is shown in Eq.~\eqref{eq_F20}. The 1-loop and 2-loop scaling functions are shown in Fig.~\ref{Fig:scaling_functions}.

\subsection{EFT predictions for the domain wall quench}\label{Ap:DW}
In this section we present the EFT predictions for the domain-wall initial condition, Eq.~(\ref{eq:dw}). As explained in the main text, the  conversion from the dynamical structure factor to the domain wall picture is a simple integration in space, Eq.~(\ref{eq:EFT_DW}). Here we present the equations employed in the fitting processes, 
\begin{align}\label{eq:EFT_eqs_dw}
   &\textbf{F}_{0,0} = \frac{1}{2}\text{Erf}(y/2)\notag\\
   &\textbf{F}_{1,0} = \frac{\chi D'^2}{\sqrt{4 \pi D^5}} \left(-\frac{4 e^{-\frac{y^2}{2}}y}{8\sqrt{\pi}} -\frac{e^{-\frac{y^2}{4}}(-6+y^2)\text{Erf}(y/2)}{16}\right)\notag\\
   &\textbf{F}_{0,1} = e^{-\frac{y^2}{4}}y(c_1 + c_2 y^2)\notag\\
   &\textbf{F}_{2,0}= \frac{\chi^2 D''^2}{24 \pi \sqrt{3}D^3}\left(\tilde{\textbf{F}}_{2,0} +\textbf{F}_{2,0}' \log t\right)\\
   &\tilde{\textbf{F}}_{2,0} = \int_0^\infty \frac{ds}{\pi} \sin (sy) s \notag\\
	&\!\! \times\left[ s^2e^{-s^2} \left(\log \frac{1}{s^2} + {\rm Ei} \left(\tfrac{2s^2}{3} \right)\right)  - \frac32 e^{-s^2/3} \right]\notag \\
 &\textbf{F}_{2,0}' = -\frac{e^{-\frac{y^2}{4}}y(y^2-6)}{16\sqrt{\pi}}.\notag
\end{align}
The parameters in the term $\textbf{F}_{0,1}$ are defined in order to absorb the numerical constant. 

Equation~(\ref{eq:EFT3}) follows from a simple manipulation of the three-point function derived in~\cite{Delacretaz:2023ypv},
\begin{equation}
\begin{split}
&f(\bar x,x,t)=\langle n(x_1,t)n(x_2,t)n(0,0)\rangle=\frac{\chi^2 D'}{
 8 \pi D^2 t} \times \\ 
 &e^{-\frac{y^2_1+y^2_2}{4}}\left(1 - \frac{\sqrt{\pi}y_1}{2}  e^{\frac{y^2_2}{4}}
 \left(\text{Erf}(y_2/2) +\text{sign}(y_1-y_2)\right)\right)+\\
&y_2 \leftrightarrow y_1,
 \end{split}
 \end{equation}
where $y_1 = x_1/\sqrt{Dt}$,  $y_2 = x_2/\sqrt{Dt}$, and $\bar x = (x_1+x_2)/2$, $x= x_1-x_2$ correspond to the centre of mass coordinates. In the domain-wall initial state the centre of mass will be summed over all positions, while the second operator is fixed at $x_2 = 0$. This corresponds to a spatial integration with respect to the center of mass coordinate,
\begin{equation}
 s_{3,\text{EFT}} = \frac{1}{\chi}  \int^0_\infty f(\bar x,x,t) d\bar x.
\end{equation}

\section{EFT for multiple diffusing densities}\label{app_mult}

\subsection{General construction of the EFT}

The EFT approach can be generalized to account for multiple continuity relations giving rise to conserved densities. Consider a thermalizing system with abelian $U(1)^N$ symmetry and couple it to background fields $A_{\mu I}^A$, where $I=1,2$ denotes the SK contour, and the index $A=1,\ldots, N$. The action will be made up of the gauge invariant combinations 
\begin{equation}
B_{\mu I}^A \equiv A_{\mu I}^A + \partial_\mu \phi_I^A\, .
\end{equation}
Imposing diagonal shift symmetry
\begin{equation}
B_{ir}^A \to B_{ir}^A + \partial_i \lambda^A(\vec x)\, , 
\end{equation}
and KMS as before, the quadratic action to leading order in derivatives is
\begin{equation}
\mathcal L^{(2)}
	= B_{0a}\cdot \chi \cdot B_{0r} + B_{ia}\cdot \sigma\cdot \left(i T B_{ia}- \dot B_{ir}\right)\, , 
\end{equation}
where $\chi_{AB},\, \sigma_{AB}$ are matrices, and dots denote matrix multiplication. Both $\chi$ and $\sigma$ have to be symmetric by time-reversal symmetry (Onsager relation). The cubic action  to leading order in derivatives is
\begin{equation}
\mathcal L^{(3)}
	= \frac12 B_{0a}\cdot \partial_A\chi \cdot B_{0r} B_{0r}^A + B_{ia}\cdot \partial_A\sigma\cdot \left(i T B_{ia}- \dot B_{ir}\right)B_{0r}^A\, .
\end{equation}
The cubic interactions arise as before as dependence of transport or thermodynamic parameters on potentials: $\partial_A\chi\equiv d\chi/d\mu^A,\,\partial_A\sigma\equiv d\sigma/d\mu^A$.
One can change variables to the density as before $\phi_r^A\to n^A \equiv \frac{\delta \mathcal L}{\delta A_{a0}^A}$. In terms of these variables, the full action up to cubic order is
\begin{align}\label{eq_L_abelian}
\mathcal L^{(2)}
	=& \, B_{0a}\cdot n + B_{ia}\cdot \sigma \cdot (iT B_{ia} - F_{0i,r}) - B_{ia}\cdot D \cdot \partial_i n\, , \notag \\ \notag 
\mathcal L^{(3)}
	=& \, -\frac12 B_{ia}\cdot \partial^A D \cdot \partial_i \left(nn_A\right)\\
	& \, + B_{ia}\cdot \partial^A \sigma \cdot (iT B_{ia} - F_{0i,r}) n_A\\ \notag
	&\, +\frac12 B_{ia} \cdot \partial^A \sigma \cdot \chi^{-1} \cdot \left(n \partial_i n_A - \partial_i n n_A\right)\, .
\end{align}
The derivatives of transport parameters are now taken with respect to densities, e.g.
\begin{equation}
\partial^A \sigma = \frac{d}{dn_A} \sigma = (\chi^{-1})^{AB} \partial_B \sigma\, ,
\end{equation}
and the diffusion matrix has been defined as 
\begin{equation}
D\cdot \chi = \sigma\, , \qquad \hbox{or} \qquad
D_A{}^B = \sigma_{AC} (\chi^{-1})^{CB}  \, .
\end{equation}
As the product of two symmetric positive matrices, $D$ can be diagonalized -- its eigenvalues correspond to the location of poles of the density two-point function $\omega = - i D_A q^2$.

While the first two lines in $\mathcal{L}^{(3)}$ \eqref{eq_L_abelian} are analogous to the $N=1$ case, the term in the third line is qualitatively new. It is a contribution to $j_{ir} = \delta \mathcal L / \delta A_{ia}$, as expected (see Eq.~\eqref{eq_j_heatcharge}). These lead to different corrections to diffusion. We study them below, in the more constrained situation of non-abelian densities.

\subsection{Non-abelian densities}

The EFT can also be straightforwardly generalized to non-abelian Lie groups \cite{Glorioso:2020loc,Akyuz:2023lsm}. We will focus here on $SU(2)$ for concreteness. Since the densities $n^A$ transform linearly (in the adjoint representation) under group action, one can implement the symmetry by making sure that they are contracted with group covariant tensors (for $SU(2)$, these are $\delta_{AB},\, \epsilon_{ABC}$). The nonlinear transformation of $\phi_a^A$ requires more attention: instead of using $B_{\mu a} = \partial_\mu \phi_a + A_{\mu a}$, one should use the Maurer-Cartan form
\begin{equation}\label{eq_B_nonabelian}
iB_{\mu a}
	=  e^{-i\phi_a^A T_A} \left(\partial_\mu + i A_\mu^B T_B\right)e^{i\phi_a^C T_C} 
\end{equation}
which also transforms in the adjoint of $SU(2)$. Here $T_A$ are the generators of the algebra. The cubic $SU(2)$ invariant action is therefore
\begin{align}\label{eq_L_nonabelian}
\mathcal L^{(2)}
	=& \, B_{0a}\cdot n + \sigma B_{ia}\cdot  (iT B_{ia} - F_{0i,r}) -  D B_{ia}\cdot \partial_i n\, , \notag\\
\mathcal L^{(3)}
	=& \, \lambda \epsilon_{ABC}B_{ia}^A n^B\left(\chi  F_{0i}^C  + \partial_i n^C \right)\, .
\end{align}
The term in the last line of \eqref{eq_L_abelian} produced the cubic term above: we have written its coefficient as $\partial_A \sigma_{BC} \equiv \chi \lambda \epsilon_{ABC}$. Note that there are also cubic terms in the first line, coming from expanding \eqref{eq_B_nonabelian}: however, one can show using the leading order equation of motion that they will not contribute to the 1-loop correction studied below.

\subsection{1-loop correction}\label{ap:B3}

Let us study the 1-loop correction to $\langle n_rn_a\rangle$. There are two contributions: the first comes from the nonlinear piece in
\begin{equation}
n_a^A
	\equiv \frac{\partial\mathcal L}{\partial A_{0r}}
	= -\sigma \nabla^2 \phi_a^A + \chi \lambda \epsilon^{ABC} \partial_i \left(\partial_i \phi_a^B n^C\right) + \cdots\, .
\end{equation}
Writing $\langle n_r^A n_a^B\rangle = \delta^{AB} \langle n_r n_a\rangle $, this leads to the following correction to $\langle n_r n_a\rangle$:
\begin{equation}
\begin{split}
	2i \chi \lambda^2 \langle n\phi\rangle(p) q_i q_j
	\int_{p'} \left[q'_j (2q'-q)_i\right] \langle n\phi\rangle(p') \langle nn\rangle(p-p')\, .
\end{split}
\end{equation}
The other contribution comes from two insertions of the cubic interaction $S^{(3)}$:
\begin{equation}
\begin{split}
&2\sigma \lambda^2  [\langle n\phi\rangle(p)]^2 q^2 q_i\times\\
&	\int_{p'} \left[q'_j(2q'-q)_i (2q-q')_j \right] \langle nn\rangle(p-p') \langle n\phi_a\rangle(p')\, .
\end{split}
\end{equation}
Summing these two diagrams one finds (in $d=1$ dimensions)
\begin{align}
&G^R_{nn}(\omega,q) = i \langle n_r n_a\rangle(\omega,q)\\ \notag 
&\simeq
	\frac{\sigma q^2}{-i\omega + D q^2}
	+ \frac{\lambda^2 T \chi^2}{D} q^2 (i\omega) \frac{\sqrt{q^2 - \frac{2i\omega}{D}}}{\left(Dq^2 - i\omega\right)^2} + \cdots \, .
\end{align}
Like before, there are several consistency checks that this result satisfies: it vanishes when $q\to 0$ (as it must by current conservation), and when $\omega\to 0$ (as it must by analyticity of static correlators). The two diagrams above do not satisfy the latter check individually, only their sum does. Our result slightly differs from one obtained in a strong noise expansion in \cite{Claeys:2021skz} -- because their result does not become analytic in the static limit, we suspect that they may have missed a contribution to the 1-loop correction.


\begin{figure*}[t]
\begin{center}
	\includegraphics[width=1.99\columnwidth]{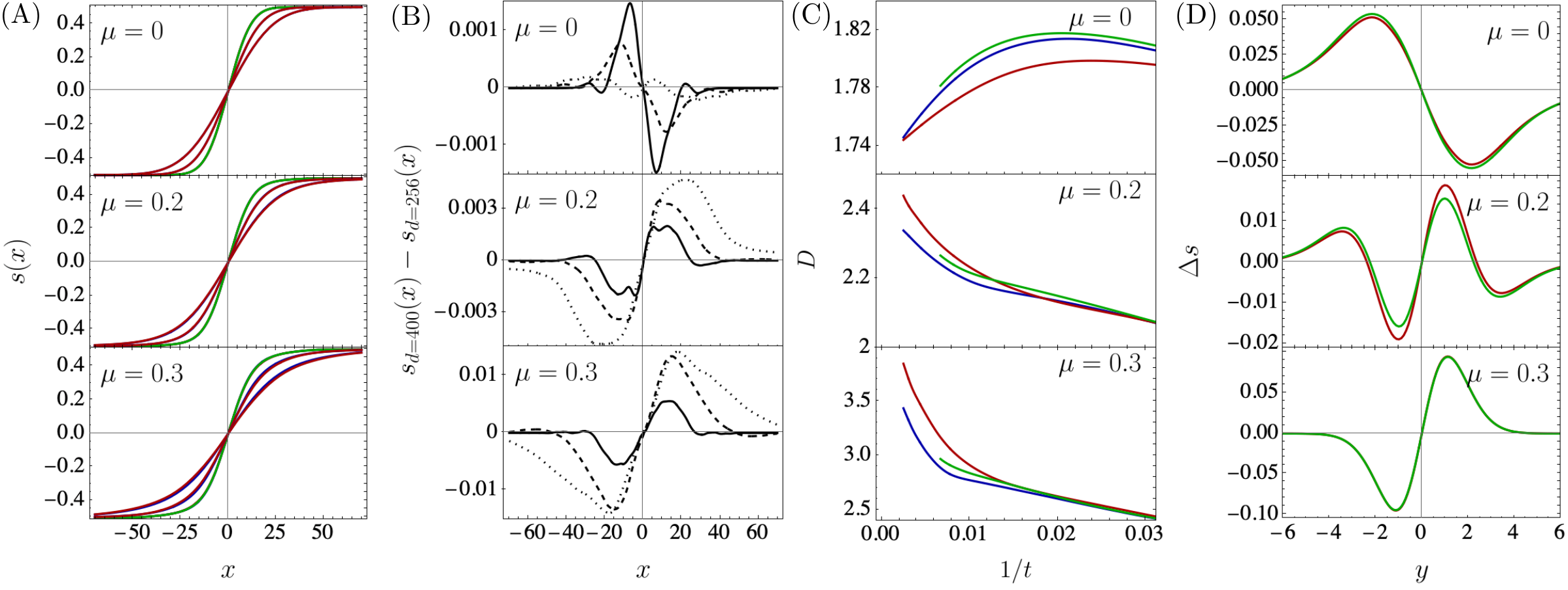}
	\caption{\label{Fig:Mag_profiles} Estimation of stability of corrections for different fillings $\mu = 0,0.2,0.3$ and bond dimensions $d = $ $256$-red, $400$-blue, $600$-green, for the staggered XXZ chain. (A): Profiles of magnetizations at 3 different times $t= 100,200,400$ (three visibly different sets of curves). Bond dimension $d = 600$ is only shown for $t =100$ the maximum integrated time is $t = 150$. A slight difference is only visible for $\mu = 0.3$. (B):  Difference between the profiles at times $t = 100,200,400$, denoted by full,dashed and dotted lines respectively. For $\mu = 0.3$ the difference is almost an order of magnitude larger than $\mu = 0, 0.2$, indicating the enhancement of simulation error at larger fillings. (C): Predictions of diffusivity by employing the fitting method (I). At $\mu = 0$ diffusivity decreases with time indicating that the system is not yet at asymptotic diffusion. At larger $\mu$ we observe a monotonic increase over time, compatible with the EFT predictions. (D): Correction to diffusion, $\Delta s = s - \textbf{F}_{0,0}$ using method II and times up to $T=150$ for $d= 256,600$. At $\mu = 0$ the difference between this plot and the main text is due to the smaller timescale used here in order to compare with $d = 600$. For $\mu = 0.2$, the  profiles are more than one order of magnitude larger than what the error is estimated from (B), suggesting that the quantitative structure of the correction is not considerably affected by errors. For $\mu = 0.3$ the profiles agree well, however deviations in (B), become significant for $t\geq 200$, and therefore, we don't fit these data in the main text.}
 \end{center}
\end{figure*}
\subsection{Fourier transformation}

We'd like to compute the fourier transform of 
\begin{align}
\delta \langle nn\rangle(\omega,q)
	&= \frac{2T}{\omega} \Im \delta G^R_{nn}(\omega,q) \\ \notag
&	=\frac{\lambda^2 (T \chi)^2}{D} q^2 \frac{\sqrt{q^2 - \frac{2i\omega}{D}}}{\left(Dq^2 - i\omega\right)^2} + \hbox{c.c.} \, .
\end{align}
Fourier transforming $\int \frac{d\omega}{2\pi} e^{-i\omega t}$ by picking up the cut as usual, one finds
\begin{equation}
\begin{split}
\delta \langle nn\rangle(t,q)
	&=   -\frac{\lambda^2 (T \chi)^2}{D^2} \frac{2}{\sqrt{\pi}}  |q| e^{-\tau/2}\times\\
	&\left[-\sqrt{\tau/2} + (\tau-1)F \left(\sqrt{\tau/2}\right)\right]\, .
\end{split}
\end{equation}
The final Fourier transform $\int \frac{dq}{2\pi} e^{iqx}$ can be performed by convoluting the Fourier transforms of both products above. The result is shown in Eq.~\eqref{eq_nn_su2}.



\section{Details on Quantum Transport}\label{Ap:QC}
In this section we (1) identify the effects of the truncation of the bond dimension in the DMPO dynamics and (2) compare different fitting approaches for the extraction 
of diffusivity. We find that even when truncation is only weakly affecting the simulation, the fact the system is not yet at the asymptotic regime can lead to different fitting results depending on the method.

\subsection{Effects of information truncation in the dynamics}
The main source of error in tensor network approaches to quantum dynamics is the truncation of information, e.g. operator entanglement in the case of density matrix evolution \cite{ZanardiPRA01}. Operator entanglement is a generated by quantum dynamics and is expected to increase linearly with time $S_{op}\propto t$, corresponding to an exponential scaling of the required classical resources (bond dimension of local tensors)~\cite{PlenioPRL09,IztokPRA08}. In practice, one sets a maximum dimension for the local tensors effectively bounding the amount of operator entanglement in the state. The effects of this truncation to the long-time dynamics in linear response quenches is an active area of research \cite{RafaelPRB18,RakPRB22}. Since there is no theory for the effects of truncation to the dynamics we simply change the bond dimension and compare the results. If there is agreement between bond dimensions we assume that the truncation is weak.

In the presence of dephasing, $\gamma$, in the system, coherences are destroyed at a timescale $t_{\gamma}\sim 1/\gamma$. This leads to a saturation of $S_{op}$. Therefore, if the bond dimension is enough to produce accurate results up to $t_{\gamma}$, it will also be accurate for $t \gg t_{\gamma}$. We confirm this for the results shown in the main text ($\gamma = 0.1$) by comparing bond dimensions $d = 128,256$ (not shown).

Coherent simulations on the other hand are much more demanding as the amount of resources increases exponentially with time. In Figure~\ref{Fig:Mag_profiles}.A we show that different bond dimensions agree well at different times at the leading order level. To get a better understanding of the accuracy we calculate the distribution differences between different bond dimensions, Figure~\ref{Fig:Mag_profiles}.B. The values of these differences will be employed to estimate the accuracy of sub-leading effects. These results show that as we deviate from half-filling the simulations become more demanding, leading to a decrease of accuracy for the same resources. While we don't fully understand this phenomenon, it's likely related to the increased strength of the non-linear corrections.

\subsection{Fitting methods}\label{sec:fit}
Following the raw data comparison, we perform a consistency check between different fitting approaches to the simulation data. The classical numerics presented in Ref.~\ref{ssec_KLS} suggest that when dealing with limited resources, making use of our knowledge of the general structure of corrections to diffusion improves the precision in fitting transport parameters. Consequently, we employ three  qualitatively different fitting methods that are designed to take an increasing amount of corrections to diffusion into account. The first two methods (I,II) perform fits on the dynamical structure factor. The third method (III) is based on Fick's first law and the total current in the system.

Method I assumes knowledge of only the leading order diffusion,
\begin{equation}\label{eq_method1_fit}
    \mathcal{S}_I(x,t) = \mathbf{F}_{0,0}= \frac{1}{2}\text{Erf}(x/\sqrt{4 \alpha_{\overline{\sigma},t} t}),
\end{equation}
where an explicit time dependence on diffusivity $ \alpha_{\overline{\sigma},t}$ is assumed in order to encapsulate the finite time corrections.

This method is simple to implement and commonly used. However it is entirely phenomenological, and, strictly speaking, incorrect -- indeed, the scaling corrections discussed in Sec.~\ref{sec_EFT} imply that the autocorrelation function does {\em not} take the form Eq.~\eqref{eq_method1_fit} at intermediate times. We nevertheless study this method to compare it to other methods that are consistent with EFT predictions. The fit is performed for each time $t$ and equilibrium magnetization $\overline{\sigma}$ on the spatial profile of magnetization $s(x)$ defined by Eq.~(\ref{eq:mag_resc}). One can then estimate diffusivity from the longest simulation time $t=T$. However, the true diffusivity is defined as $D = \lim_{t\rightarrow \infty}\alpha_{\overline{\sigma},t}$. In principle, it is possible to extrapolate diffusivity with some appropriate function of inverse time. However, in simulations which have increasing errors with time, such extrapolations can capture artifacts, so we will not be performing them.  Once diffusivity $D(\overline{\sigma})$ is extracted, the non-linear corrections are either calculated by discrete  derivatives, or  analytically, by first performing a low order polynomial fit,
\begin{equation}\label{eq:Diff_trial}
    D(\overline{\sigma}) = \sum^M_{i=0}b_{i}(\overline{\sigma})^{2 i},
\end{equation}
where only even powers appear due to the particle-hole symmetry in the system. In Figure~\ref{Fig:Mag_profiles}.C we show how this method performs on the staggered field simulations for different bond-dimensions. Compared to methods II and III, I gives a time-dependent illustration of the systems behaviour, by showing how the asymptotic limit is approached.

Method II, which is explained in the main text is based on a  full fit of the hypersurface $s(\overline{\sigma},x,t)$. This is done by either employing the leading order diffusion or more elaborate assumption for the fitting function, Eq.~(\ref{eq:EFT_DW}) based on our knowledge of the leading corrections to diffusion. We use the same trial function for diffusivity as before, Eq.~(\ref{eq:Diff_trial}) and similar trial functions for the linear corrections, $c_{1/2} = \sum^M_{i=0}b^{1/2}_{i}(\overline{\sigma})^{2 i}$. We note that  $c_{1/2}$ are taken constant in the coherent simulations since the maximum reliable chemical potential is $\approx 0.25$ and the dependence of the constants with chemical potential is weak. Figure~\ref{Fig:Mag_profiles}.D shows the correction to the leading order diffusive profile at $T= 150$. We observe that for sufficiently low $\mu \leq 0.25$ the correction is more than one order of magnitude stronger than the bond dimension difference (Figure~\ref{Fig:Mag_profiles}.B), indicative of a quantitatively accurate result. For the incoherent simulation where we fit $\mu = 0-1$ it is important to allow for the constants to depend on $\bar \sigma$.

The third method (III) studies the relaxation of the total current $J=\int dx \, j(x)$. On general grounds, the constitutive relation for the current density is
\begin{equation}\label{eq_Fick}
    j = -D(n) \partial_x n  + \hbox{higher derivatives}.
\end{equation}
We will ignore the higher derivative terms for now, and come back to them below. The first term is responsible for all higher-loop corrections to the dynamic structure factor at leading order in gradients, $\textbf{F}_{\ell,0}$ (the first two $\ell=1,2$, coming from $D',D''$, were computed in Sec.~\ref{sec_EFT}). All of these corrections do not contribute to correlators of the total current: defining $C(n)$ such that $C'(n) = D(n)$ and integrating from $x=1$ to $x=L$, we have
\begin{equation}\label{eq_totalcurrent}
\begin{split}
J &\simeq C(n(L))- C(n(1)) \\
  &\simeq D(\bar \sigma) \left(s(L)-s(1)\right),
\end{split}
\end{equation}
where in the second line we expanded the density around the equilibrium magnetization $n(x)=\bar\sigma + s(x)$, and $D(\bar\sigma)$ is approximated by Eq.~(\ref{eq:Diff_trial}). It is justified to drop higher order terms in $s(x)$ (which could otherwise lead to fluctuating corrections), because the dynamics has not affected the magnetization sufficiently far from the domain wall position at any time. This is true since for all times integrated, the Lieb-Robinson's light-cone has not reached the system's boundary, meaning that $s(L) = -s(1) \simeq \chi(\bar\sigma) \delta$. Let us now turn to the higher-derivative terms in Eq.~\eqref{eq_Fick}.  Because of the argument above, $\partial^n_{x} s(x) = 0$ at the boundaries $x=1,L$ so that other contributions to the current that are total derivatives vanish. This includes linear higher derivative terms $j\supset \partial_x^{2n+1} s$, which otherwise would have lead to corrections of the form $\textbf{F}_{0,n}$. This also includes certain nonlinear higher derivative terms: for example, all terms involving two diffusive fluctuations are total derivatives $s\partial_x^{2n+1}s = \partial_x (\ldots)$, so that 1-loop corrections with any number of derivatives $\textbf{F}_{1,n}$ do not contribute. The first correction to \eqref{eq_totalcurrent} comes from the leading EFT operator that is parity-odd, and not a total derivative. For the case of a single diffusive density, this is $j\supset s^2 \partial_x^3 s$ \cite{PhysRevB.73.035113}. This will lead to a $\ell=2$ loop correction at order $n=2$ in derivatives ($\textbf F_{2,2}$) to the current decay $J(t)$ scaling as $1/t^{n+\ell/2} = 1/t^3$ at late times. While a precise exponent is difficult to extract from the numerics, we observe fast decay of the total current, consistent with the observation that many of the leading EFT corrections vanish in this observable (fast polynomial decay due to high order hydrodynamic tails is known to be difficult to observe quantitatively \cite{Glorioso:2020loc}). Because the convergence to the late time value of the current $\lim_{t\to \infty} J(t)$ is therefore fast, there is no need to extrapolate in time to obtain accurate results.  We note that this method becomes less powerful in the presence of multiple conserved charges, since there is a larger number of terms which are not total derivatives, see Eq.~\eqref{eq_j_heatcharge}. In this case, the decay of the current is due to a 1-loop correction $\textbf{F}_{1,0}$, scaling as $1/t^{n+\ell/2} = 1/\sqrt{t}$.

\begin{figure}[t!]
\begin{center}
	\includegraphics[width=0.9\columnwidth]{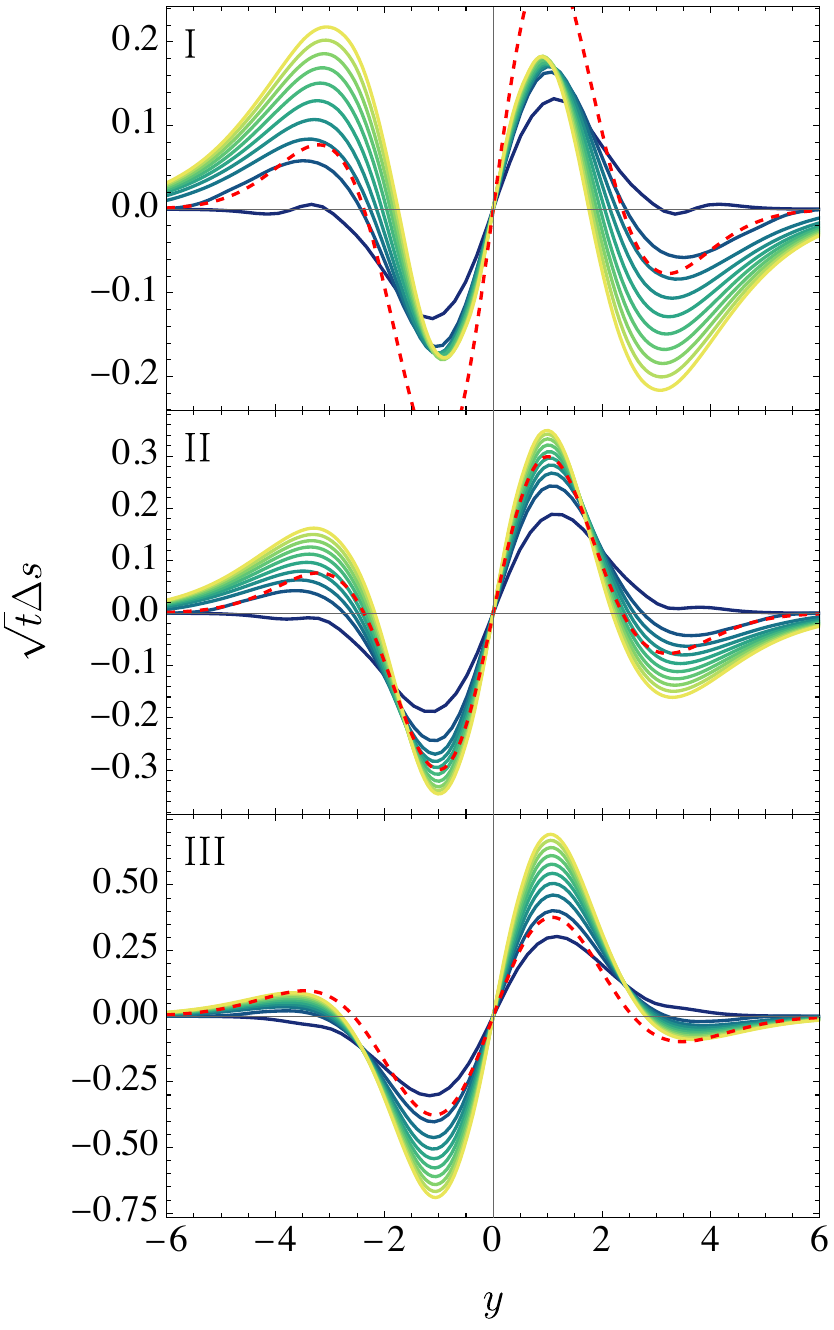}
	\caption{\label{Fig:Cor_comp} Sub-leading correction estimation for diffusivity fits using different fitting methods for $\mu = 0.2$. Faster increasing diffusivity results to suppressed edge corrections, while slower increasing diffusivity leads to suppressed corrections around $y= 0$. The analytical result (red dashed line) denotes the EFT result \textbf{F}$_{1,0}$, which relies only on the fitted diffusivity $D(\overline{\sigma})$. } 
 \end{center}
\end{figure}

The diffusivity $D(\bar\sigma)$ is therefore obtained from the late time current using \eqref{eq_totalcurrent}, and fit as a function of magnetization using \eqref{eq:Diff_trial}, limiting ourselves to $M=2$. The parameters $b_0,b_1,b_2$ are estimated by minimizing the distance between the measured currents at different fillings and Eq.~\eqref{eq_Fick}. To avoid over-parametrization, we fit a number of equilibrium magnetization much larger than the number of free parameters (3 in this case).

The sub-leading correction is qualitatively similar for fitting methods II and III despite the small variations in diffusivity, Figure~\ref{Fig:Cor_comp}, however method I clearly underestimates diffusivity, leading to larger deviation in the corrections.


\section{Non-linear response}\label{sec_nonlin}\label{ap:NL}
In the previous sections, we explored the effects of scaling corrections in the DSF which is a two-point correlation function. While the effect of fluctuations on the DSF can be important, it is always subleading at long times. Here, we go one step further and explore many-body correlations which would vanish in the absence of nonlinearities in hydrodynamics.  In particular we explore the late-time behavior of the observable,
\begin{equation}
s_3(x,t)=\frac{\langle \sigma^z\left(L/2,t\right)\sigma^z\left(L/2+x,t\right)\rangle_c}{  | 2\text{tr} \left( \sigma^z_{1} \rho(t=0) \right)|}, 
\end{equation}
where the average is performed on the domain wall state defined by Eq.~(\ref{eq:dw}). As we illustrate in Appendix~\ref{Ap:DW}, this observable is, up to a spatial integration, a special case of the density three-point function. Higher-point functions generalize full counting statistics in that they allow for operator insertions at multiple times, and can be obtained from the EFT of diffusion~\cite{Delacretaz:2023ypv}.  According to the EFT, the asymptotic behavior of $s_3(x,t)$ in diffusive systems with a single conserved charge is given by a universal scaling function
\begin{equation}\label{eq:EFT3}
    s_{3,\text{EFT}}(y,t) =\frac{1}{\sqrt{t}}\frac{D'\chi}{8\sqrt{D^3 \pi}} \left(e^{-y^2/4}+\text{Erf}(y/2)-1\right),
\end{equation}
where $y= x/\sqrt{D t}$. As expected for multi-body functions in linear response,  $s_{3,\text{EFT}}(y,t)$ is vanishes as $t\rightarrow \infty$. It also vanishes in systems which are particle-hole symmetric ($D'=0$).

\begin{figure}
	\centering
\includegraphics[width=0.95\columnwidth]{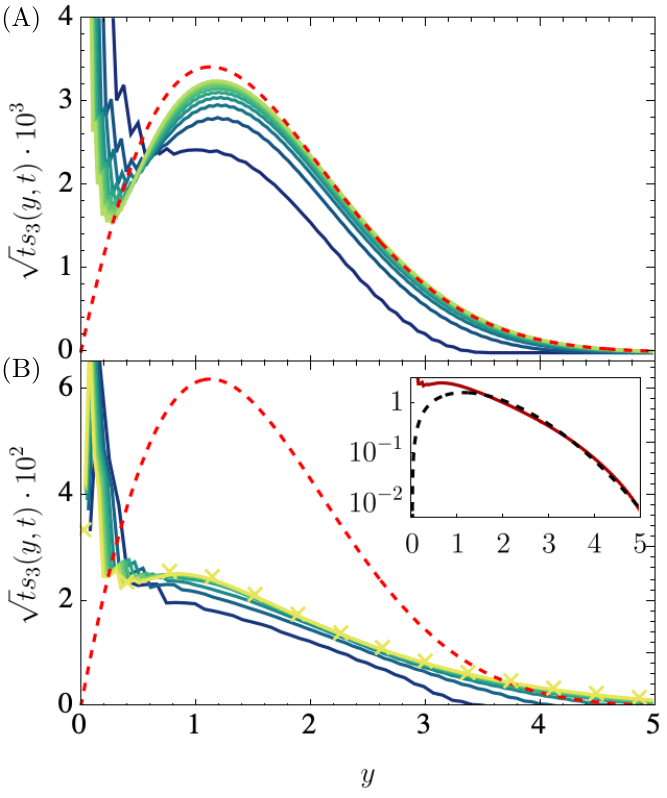}\caption{\label{Fig:5} Three-point correlation at linear response.  (A) Noisy system with bond dimension $d=256$, $\gamma = 0.1$ at $\mu = 0.65$. (B) Staggered field with bond dimension $d= 600$,  $g = 0.4$ at $\mu = 0.23$. Different coloured curves denote different times $t \in\{40, 400\}$ for (A) and $t \in\{40, 200\}$ for (B)  with smaller times corresponding to darker colours. Red dashed lines corresponds to EFT predictions with diffusivity estimated by method III. Crosses in (B) show bond dimension $d = 400$ results at $t= 200$. Inset: The tail of the distribution at $t=200$ against the EFT function (black dashed) where $(D,D')$ are taken as fitting parameters. }
\end{figure}

As shown in Figure~\ref{Fig:5}, for both dephasing and staggered perturbations, $s_3$ scales according to the EFT prediction, $1/\sqrt{t}$. In the presence of dephasing, we observe a precise late time agreement of the correction profile to $s_{3,\text{EFT}}$, which verifies the validity of the EFT prediction. No fitting parameter was used in this test, as $D$ and $D'$ had already been obtained from the linear response analysis. In the case of staggered field, we find that while the shape of the profile is qualitatively similar, there is a quantitative deviation from the EFT prediction (independent of the diffusivity fitting method). The similarity between the EFT profile \eqref{eq:EFT3} and the numerical data can be illustrated by performing a fit of $s_{3,\text{EFT}}$ for $y>1$, with now $D$ and $D'$ taken to be independent fitting parameters. The result is illustrated in the inset of Figure~\ref{Fig:5}.B, the two profiles agree well for $(D\sim 1.85 D_{\text{III}},D'\sim 0.36 D'_{\text{III}})$. This agreement suggests that the staggered profile at $y\gg 1$ has the same functional form as the EFT prediction and features a Gaussian tail, despite the apparent disagreement with the values of $D$ and $D'$ obtained from linear response.

While the discrepancy between the EFT prediction and quantum dynamics in staggered field simulations remains unclear to us, we have verified that this disagreement is not due to simulation errors. Additionally, the good convergence with time suggests that subleading corrections are not at play. Taking into account these observations leaves us with several possibilities: (i) The diffusivity fits of the previous section may not be accurate for the staggered field simulations. This could be an artifact of the short accessible time scale or of large diffusivity fluctuations with magnetization. We have extensively checked for these artifacts and we did not find any evidence of them in the data. (ii) This system may simply have a very long local equilibration time, $\tau_{\rm eq}\gtrsim 200$, so that our numerics never probes the truly asymptotic regime controlled by the EFT. This could e.g. arise from the presence of additional long-lived degrees of freedom protected by approximate symmetries, and could be related to the integrability of the Floquet-XXZ chain, or to other prethermalization mechanisms~\cite{Abanin17CMP}. (iii) Finally, the EFT may fail to capture even the asymptotic dynamics of coherent many-body Floquet systems. While we do not have a particular reason to believe that the EFT should fail, there is no proof that it must emerge in general. We leave this interesting possibility, as well as further exploration of these phenomena, for future work.

\end{document}